\documentclass[11pt, a4paper]{article}
\usepackage[utf8]{inputenc}
\usepackage{authblk}
\usepackage{upgreek}
\usepackage{graphicx}
\usepackage{amsmath}
\usepackage[margin=3cm]{geometry}
\usepackage{makecell}
\usepackage[shortlabels]{enumitem}
\usepackage{amssymb}
\usepackage{subcaption}
\usepackage[rightcaption]{sidecap}
\usepackage{longtable}
\usepackage{xcolor}
\usepackage{hyperref}
\usepackage{svrsymbols}
\usepackage{marvosym}
\usepackage{comment}
\usepackage[font=small, labelfont=bf, justification=raggedright, format=hang]{caption}
\usepackage{floatrow}
\usepackage{wrapfig}
\usepackage{units}
\usepackage[T1]{fontenc}
\usepackage{multirow, bigdelim, bigstrut}
\usepackage[toc]{appendix}
\usepackage{placeins}
\usepackage{rotating}
\usepackage{pdflscape} 
\usepackage{bbm} 
\usepackage[autostyle]{csquotes}

\usepackage{physics}
\usepackage{mathtools}
\usepackage{mathrsfs}

\font\titlefont=cmr12 at 16pt
\font\authfont=cmr12 at 12pt
\newcommand\afffont{\fontfamily{cmr}\fontsize{10}{12}\itshape}
\title{{\titlefont Development and characterisation of high-resolution microcalorimeter detectors for the ECHo-100k experiment}}

\author[1, $\spinup$ ,*]{\authfont F. Mantegazzini}
\author[1, $\spin$,*]{\authfont N. Kovac}
\author[1]{\authfont C. Enss}
\author[1]{\authfont A. Fleischmann}
\author[1,$\spindown$]{\authfont M. Griedel}
\author[1]{\authfont L. Gastaldo}

\affil[1]{\afffont Kirchhoff Institute for Physics, Heidelberg University}
\affil[$\spinup$]{\textit{\normalsize Current affiliation: Center for Sensors and Devices, Fondazione Bruno Kessler}}
\affil[$\spin$]{\textit{\normalsize Current affiliation: Tritium Laboratory Karlsruhe - Institute for Astroparticle Physics, Karlsruhe Institute of Technology}}
\affil[$\spindown$]{\textit{\normalsize Current affiliation: Institute for Quantum Materials and Technologies, Karlsruhe Institute of Technology}}
\affil[*]{\textit{\normalsize Corresponding authors}}

\date{}

\begin{document}

\maketitle

\begin{abstract}

\noindent  The goal of the ECHo experiment is a direct determination of the absolute scale of the neutrino mass by the analysis of the end-point region of the $^{163}\mathrm{Ho}$ electron capture (EC) spectrum. The results of the first phase of the experiment, ECHo-1k, have paved the way for the current phase, ECHo-100k, which aims at a sensitivity below $2 \, \mathrm{eV}$ on the effective electron neutrino mass. In order to reach this goal, a new generation of high-resolution magnetic microcalorimeters with embedded $^{163}$Ho have been developed and characterised. The design has been optimised to meet all the challenging requirements of the ECHo-100k experimental phase, such as excellent energy resolution, wafer scale implantation and multi-chip operation with multiplexing read-out.
We present the optimisation studies, the final design of the detector array and the first characterisation studies. The obtained results demonstrate that the detectors fully match and even surpass the requirements for the current experimental phase, ECHo-100k.

\end{abstract}

\section{Introduction}

The ECHo experiment aims to investigate the effective mass of the electron neutrino on a sub-electronvolt level, by applying a model-independent method. 
Specifically, the strategy is based on the analysis of the calorimetric electron capture (EC) spectrum of $^{163}$Ho \cite{DeRujula_Lusignoli} \cite{ECHoGeneral}. During an EC process, an electron from an inner shell of the $^{163}$Ho atom is captured by the nucleus and an electron neutrino is emitted. Consequently, the excited daughter atom relaxes to the ground state, emitting cascades of electrons and, in a small fraction of cases, also X-rays.
The minimum energy required to create a neutrino corresponds to its mass and, therefore, the maximum energy in the measured de-excitation EC energy spectrum is reduced by that amount. 
As a result, the spectral shape at the end-point is affected by the finite neutrino mass. 

The current best limit on the effective electron antineutrino mass is $0.8 \, \mathrm{eV}$ 90\% at C.L., obtained by the KATRIN experiment \cite{KATRIN2022}, while the current best limit on the effective electron neutrino mass is $150 \, \mathrm{eV}$ at 90\% C.L. and it was obtained by the ECHo collaboration with a preliminary calorimetric measurement of the $^{163}$Ho EC spectrum \cite{ECHo_spectrum_2019}. 
Another experiment that is exploiting cryogenic microcalorimeters to measure the $^{163}$Ho EC spectrum in order to probe the electron neutrino mass is HOLMES \cite{Holmes2023}. 

During the first phase of the ECHo experiment, ECHo-1k, a dedicated 72-pixel detector array of metallic magnetic calorimeters (MMCs) \cite{Fle2005} has been developed and characterised \cite{ECHo-1k} and a high-statistics spectrum with about $10^{8}$ $^{163}$Ho EC events in total has been acquired. 
According to the sensitivity calculations based on the most recent theoretical description of the $^{163}$Ho EC spectrum \cite{Brass_HoSpectrum_2020, ECHoGeneral}, with the on-going data analysis a new limit below $20 \, \mathrm{eV}$ on the effective electron neutrino mass will be set. 
The results from the ECHo-1k phase have paved the way for the current phase, ECHo-100k, with the goal to collect a total statistics of about $10^{13}$ events, allowing for a sensitivity below $2 \, \mathrm{eV}$. 

In order to meet the new requirements for the ECHo-100k experiment, focused optimisation studies have been performed and a new detector array has been designed, fabricated and characterised, showing excellent performance. 

\section{Detector requirements}
\label{SEC:requirements}

The detector technology of choice for the ECHo experiment must satisfy four main requirements, namely high energy resolution, ability to embed a sufficient amount of $^{163}$Ho atoms, fast detector response and suitability for multiplexed read-out.
A high energy resolution is a fundamental requisite to ensure a precise measurement of the end-point region of the $^{163}$Ho spectrum, avoiding possible smearing which would compromise the sensitivity of the experiment. 
The $^{163}$Ho atoms must be completely enclosed inside the detector to allow for a calorimetric approach of the measurement. 
A fast response minimises the unresolved pile-up fraction due to the high source activity and, finally, a multiplexed read-out becomes essential when the number of detectors operated in parallel is larger than $\sim 100$. 

The detector technology chosen by the ECHo experiment is based on low temperature metallic magnetic calorimeters (MMCs) \cite{Fle2005}. The working principle of MMCs is based on a paramagnetic temperature sensor thermally coupled to a gold absorber. The temperature raise that follows an energy deposition in the absorber is translated into a change of magnetisation that can be read out by a superconducting pick-up coil placed underneath the sensor. Finally, a Superconducting Quantum Interference Device (SQUID) is used to convert the flux change in the pick-up coil into a voltage signal.

MMCs developed for X-ray detection have shown an optimal energy resolution - up to $1.6 \, \mathrm{eV}$ FWHM at $6 \, \mathrm{keV}$ -, fast response - reaching a rise-time of the signal of about $90 \, \mathrm{ns}$ -, good linearity and, thus, reliable calibration \cite{Fle2005} \cite{Kem2018}. Furthermore, it has been already demonstrated that MMCs can be tailored to the specific requirements of the ECHo experiment \cite{Gastaldo_NIMA} \cite{Hassel2016} \cite{ECHo_spectrum_2019} \cite{ECHo-1k} \cite{LTD_proceedings_2021}. A dedicated $^{163}$Ho ion-implantation and post-processing procedures have been developed and successfully tested \cite{ECHo-1k}.
The implantation procedure includes also the subsequent deposition of a second absorber layer to enclose the $^{163}$Ho source. 
Additionally, a microwave SQUID multiplexing read-out for MMCs has been recently developed and demonstrated \cite{MUX_2018} \cite{MUX_2019} \cite{Richter2021} \cite{Ahrens2022}.

In the first phase of the ECHo experiment, ECHo-1k, about $10^{8}$ EC events from $^{163}$Ho source the have been acquired using about 60 MMC detectors with an activity of about 1 Bq/pixel\footnote{1 Bq of activity corresponds to about $2\cdot 10^{11}$ implanted $^{163}\mathrm{Ho}$ ions.} and are currently being analysed, with the goal of reaching a sensitivity on the effective electron neutrino mass below $20 \, \mathrm{eV}$ 90\% C.L.. 
The current phase, ECHo-100k, aims to improve the sensitivity by one order of magnitude, by reaching at least $2 \, \mathrm{eV}$ 90\% C.L..
Specifically, a significant increase of the statistics is required, with a total of $10^{13}$ EC events, while still maintaining a reasonable acquisition time of a few years. These requirements translate into a total $^{163}$Ho activity of 100\,kBq. The amount of $^{163}$Ho activity per MMC pixel is determined by the allowed pile-up fraction\footnote{The ECHo experiment aims to push the external background below a level of about $10^{-5}$\,counts/eV/det/day \cite{ECHoGeneral}. Therefore, the allowed unresolved pile-up fraction is set to $10^{-5}$.} and is about 10\,Bq, which translates into a total required number of MMC pixels of $10^4$ which need to be operated in parallel.

The main upgrades from ECHo-1k to ECHo-100k are summarised in table \ref{TAB:ECHo-100k_requirements}. 

\begin{table}[h!]
\begin{center}
\begin{tabular}{ |c|c|c| } 
\hline
& \textbf{ECHo-1k} & \textbf{ECHo-100k} \\
\hline
Activity & $a \approx 1 \, \mathrm{Bq/pixel}$ & $a \approx 10 \, \mathrm{Bq/pixel}$  \\
Detector energy resolution & $\Updelta E_{\mathrm{FWHM}} \leq 10 \, \mathrm{eV}$ & $\mathrm{\Delta} E_{\mathrm{FWHM}} \leq 5 \, \mathrm{eV}$  \\
Number of pixels & $N = 57$ & $N = 12000$ \\
Read-out technology & dc-SQUID read-out & $\upmu$MUXing* \\
\hline
\end{tabular}
\end{center}
\footnotesize{* Microwave SQUID multiplexing}
\caption{Overview of the experimental requirements for the ECHo-100k phase in comparison with the ECHo-1k phase.}
\label{TAB:ECHo-100k_requirements}
\end{table}

\section{Detector concept}
\label{SEC:detector}

\subsection{Design}
\label{SUBSEC:design}

The ECHo-100k array consists of a chip with a size of $13 \, \mathrm{mm} \times 5 \, \mathrm{mm}$ hosting 64 MMC pixels, corresponding to 32 read-out channels. 
Each pixel consists of a niobium superconducting meander-shaped pick-up coil, a Ag:Er paramagnetic sensor placed on top and connected to an on-chip gold thermal bath and two absorber layers enclosing the $^{163}$Ho source, similarly to the design of the detectors used for the ECHo-1k experimental phase \cite{ECHo-1k}.

The first absorber layer stands on three gold cylindric pillars with a diameter of about $16 \, \mathrm{\upmu m}$ so that the contact area with the sensor is reduced, preventing athermal phonons to travel through the sensor and release the energy in the substrate. The $^{163}$Ho atoms are ion-implanted in a dedicated host material layer, with a thickness in the order of $100 \, \mathrm{nm}$, that is deposited on top of the first absorber layer. After the implantation process, a second host material layer with similar thickness is deposited before structuring the second absorber layer on top, thus fully enclosing the $^{163}$Ho source.
The chosen host material is silver, in order to minimise the heat capacity contribution. In fact, silver implanted with holmium ions shows a smaller heat capacity in the temperature range around 20\,mK if compared to gold implanted with holmium ions \cite{Ho_Au_HC}. 

The necessary weak magnetic field that polarises the spins in the sensor is provided by a persistent current that is injected in the superconducting meanders exploiting a dedicated heater switch based on a resistive Au:Pd element \cite{ECHo-1k}.

\begin{figure}[h!] 
    \centering
    \includegraphics[width=0.5\textwidth]{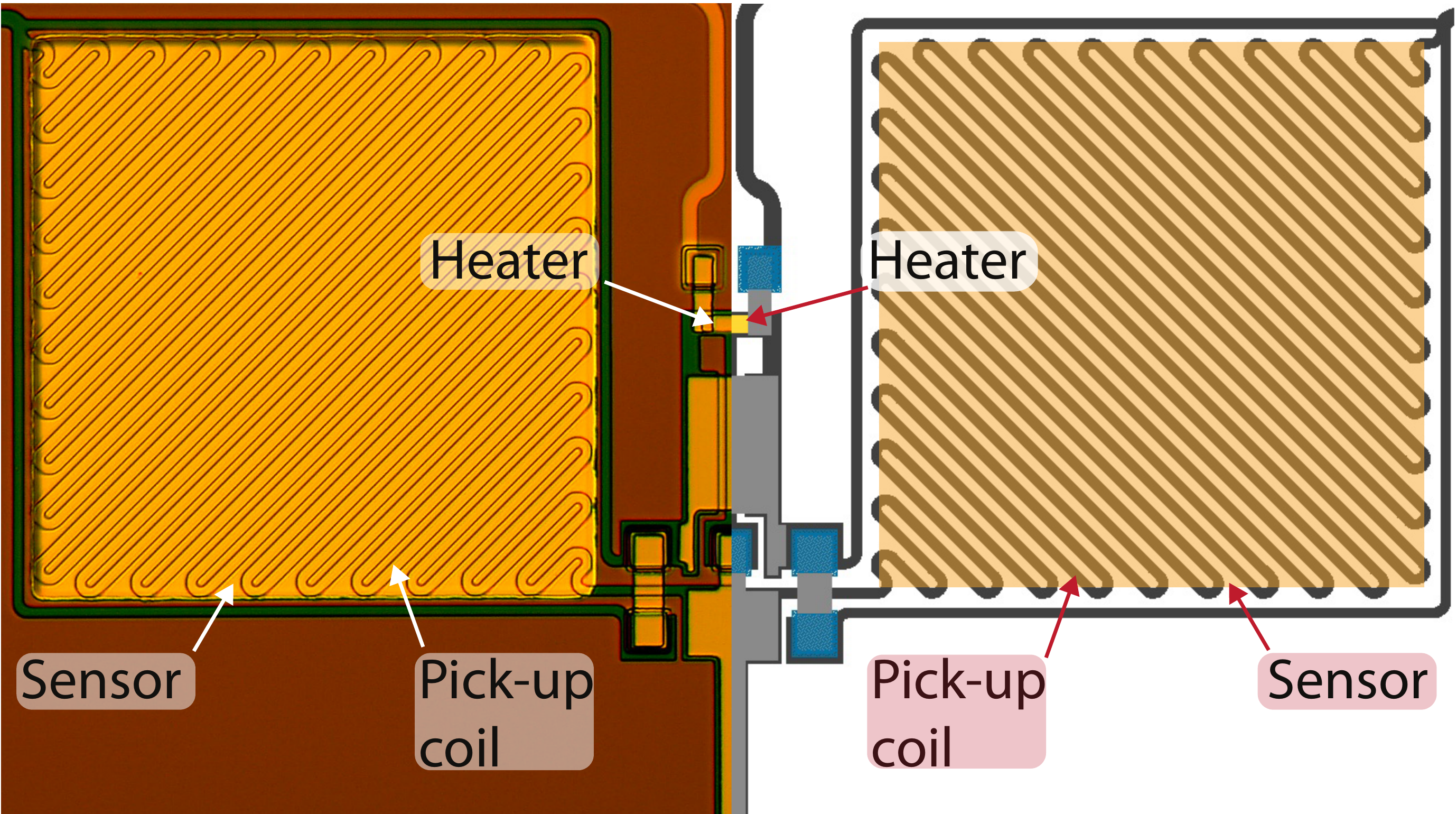}
    \caption{Microscope photo (left half of the picture) and schematic design (right half of the same picture) of a detector channel of the ECHo-100k design, consisting of two MMC pixels. The meander-shaped pick-up coil and the sensor are present, while the absorber layers are not deposited yet. The normal conducting element that functions as a heater switch is highlighted.} 
    \label{FIG:single_pixel}
\end{figure}

The two meander-shaped pick-up coils are connected in parallel to the input coil of the same SQUID read-out device, forming a first order gradiometer \cite{Fle2005}.
As a result, the signal polarity of the two pixels is opposite and common changes of temperature (e.g.~small changes of the chip temperature) cancel out. 
Figure \ref{FIG:single_pixel} shows the layout of a single detector channel (microscope photo on the left part and design on the right part).

\begin{figure}[h!] 
    \centering
    \includegraphics[width=\textwidth]{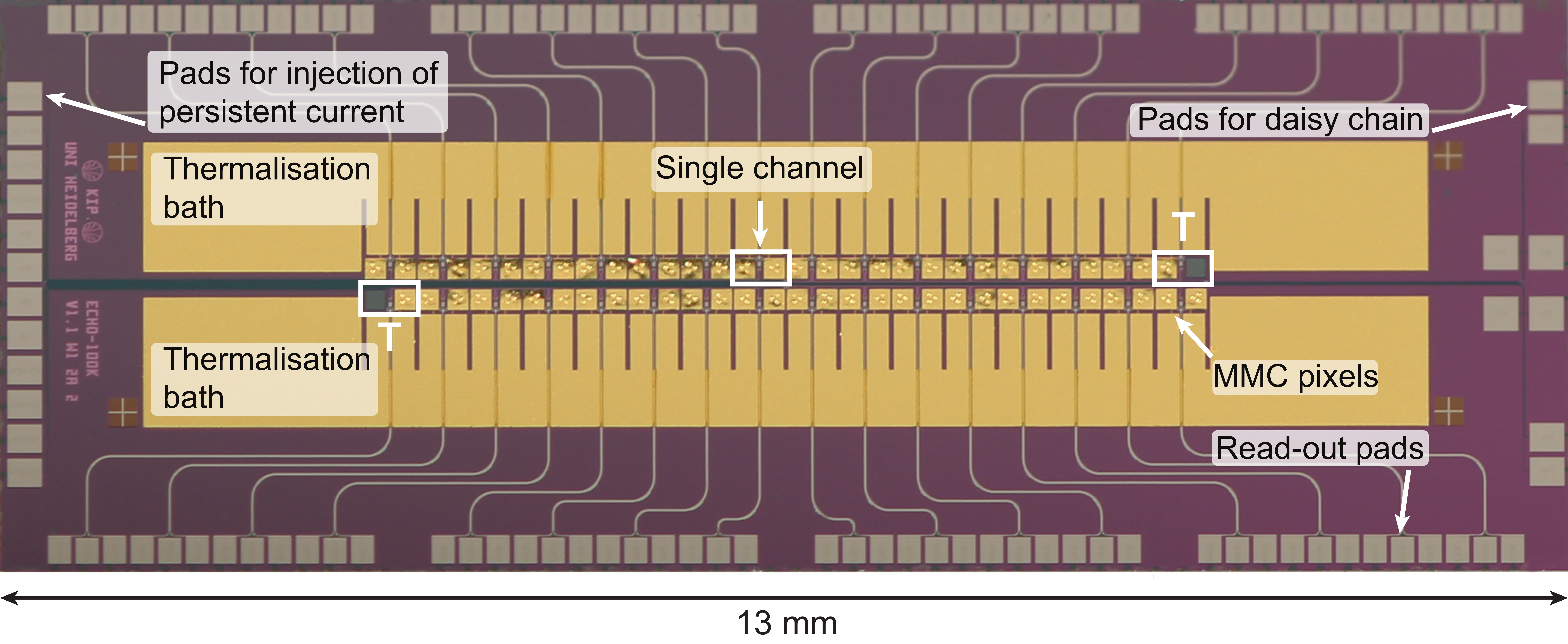}
    \caption{Photograph of the 64-pixel detector array chip developed and produced for the ECHo-100k phase. The two temperature monitoring channels are marked with the letter T. The pads dedicated to read-out and to the injection of the persistent current in the pick-up coils, the thermal baths and the two temperature monitoring channels are highlighted.} 
    \label{FIG:ECHo-100k_chip}
\end{figure}

Two detector channels of the array are characterised by a non-gradiometric layout, where the sensor is present only on one side of the double meander coil, making them sensitive to temperature changes. These two channels are dedicated to temperature monitoring. 
Figure \ref{FIG:ECHo-100k_chip} shows a photograph of the detector array chip. In the centre, the array comprised of 64 MMC pixels, i.e.~32 detector channels, is visible. At the two opposite sides of the array, the two non-gradiometric temperature monitoring channels are present. The large gold areas serve as an on-chip thermalisation baths and they are connected in series using microfabricated air bridges, as explained in the following. The pads at the periphery of the chip are dedicated to the read-out and to the injection of the persistent current into the pick-up coils. 

The optimisation studies performed to design a suitable detector array for the ECHo-100k phase have been focused on four main objective: a) improving the energy resolution, b) suppressing energy losses, c) improving the on-chip thermalisation, d) upgrading to multichip operation.


\subsubsection*{Energy resolution}

The energy resolution $\Delta E_\mathrm{FWHM}$ of microcalorimeters scales with the operational temperature $T$ and with the square root of the total heat capacity $C$ \cite{Fle2005}. The volume of the gold absorber, inside which the $^{163}$Ho source is embedded, and the sensor affect the detector heat capacity. 
During the implantation phase, the host material needs to be co-sputtered to avoid saturation effects \cite{Gamer2017}. 
In the detectors developed for the ECHo experiment, decreasing the implantation area for the same $^{163}$Ho activity would increase the amount of co-sputtered host material, thus increasing also the total heat capacity. 
Furthermore, the area of the absorber cannot be set arbitrarily small without compromising the alignment for the post-processing step for the fabrication of the second absorber layer after the $^{163}$Ho implantation. 
Thus, the best way to minimise the absorber volume is to reduce the absorber thickness.
Dedicated simulations have been set up with the goal of minimising the absorber thickness, while ensuring enough stopping power. 
In fact, for the purpose of the ECHo experiment it is crucial to keep high quantum efficiency in the full spectrum energy range in order not to distort the shape of the spectrum, especially in the end-point region. 

\begin{figure}[h!] 
    \centering
    \begin{subfigure}[b]{0.49\textwidth}
        \centering
        \includegraphics[trim=12 0 12 0,clip,width=0.99\linewidth]{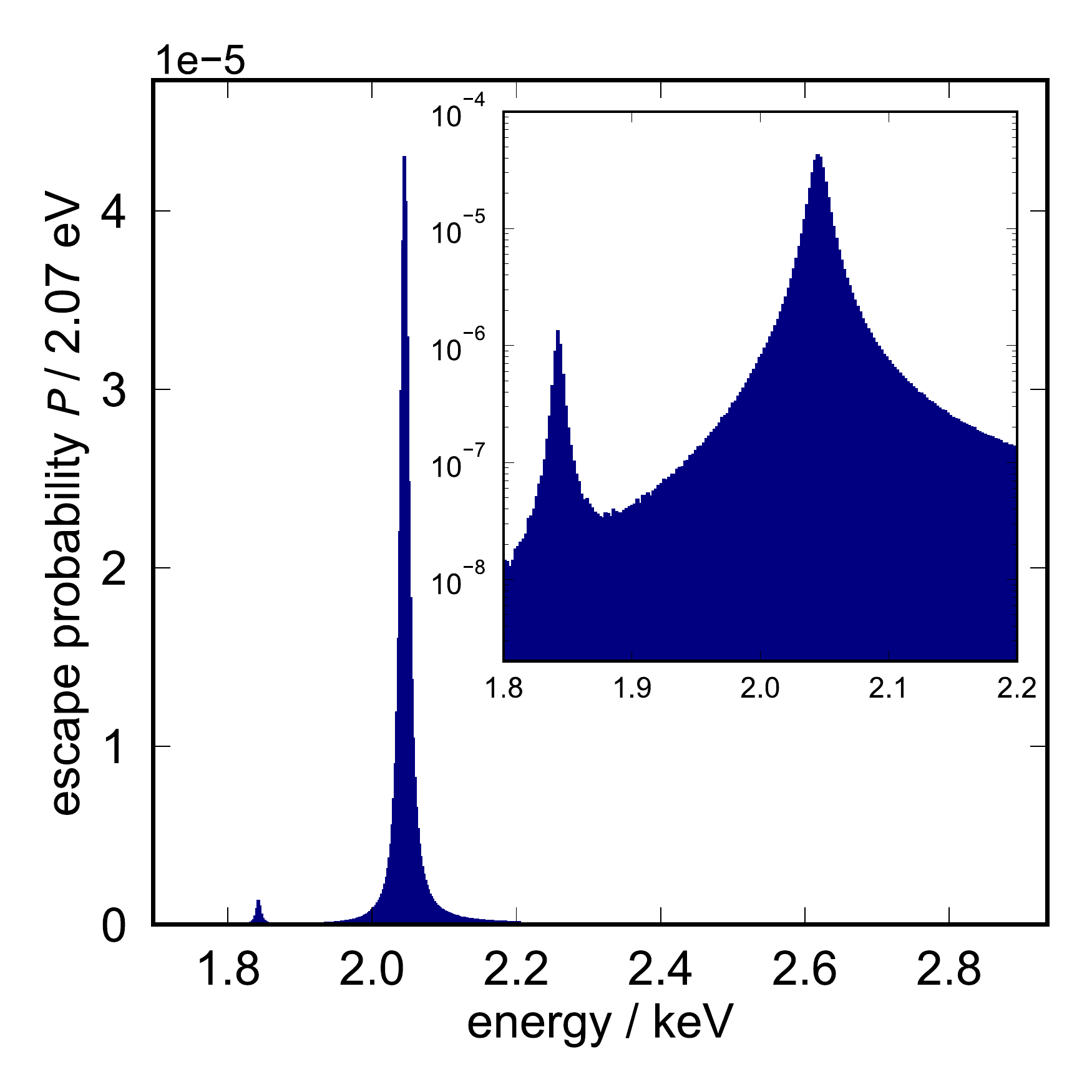}
        \caption{} \label{SUBFIG:P_surv_all_range}
    \end{subfigure}
    \hfill
    \begin{subfigure}[b]{0.49\textwidth} 
        \centering
        \includegraphics[trim=12 0 12 0,clip,width=0.99\linewidth]{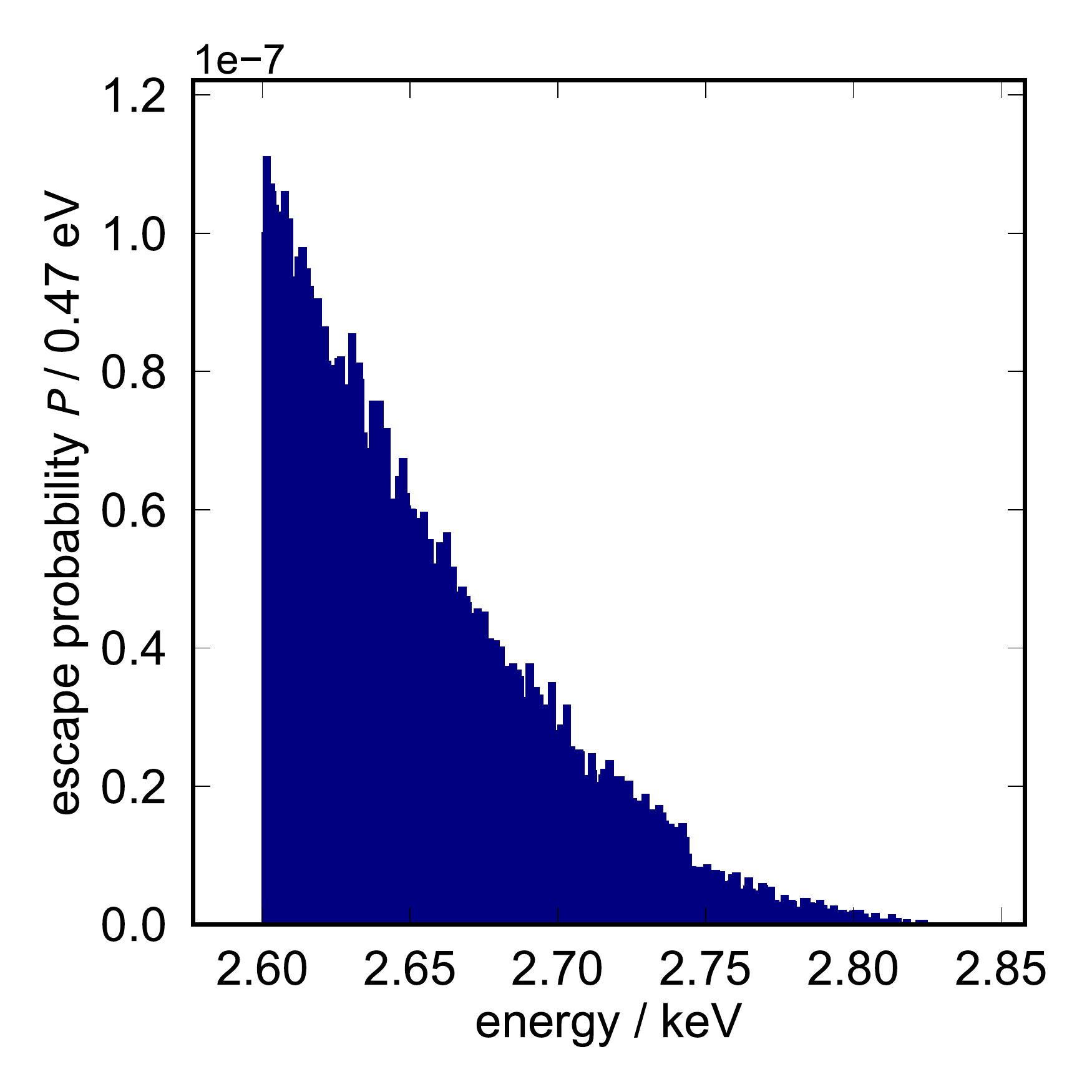}
        \caption{} \label{SUBFIG:P_surv_endpoint_range}
    \end{subfigure}

    \caption{Escape probability histograms calculated with Monte Carlo simulations with $10^9$ events assuming an absorber thickness of $3.0 \, \mathrm{\upmu m}$. In \textbf{a)} the energy range is between $1.800 \, \mathrm{keV}$ and $Q_{\mathrm{EC}} = 2.833 \, \mathrm{keV}$ and in the inset a magnification in the energy region of the MI and MII resonances is shown in logarithmic scale. In \textbf{b)} the energy range is between $2.600 \, \mathrm{keV}$ and $Q_{\mathrm{EC}} = 2.833 \, \mathrm{keV}$. The discontinuity visible at $2.75 \, \mathrm{keV}$ is due to the absorption edge corresponding to the binding energy of the M3-shell of gold \cite{NIS00}.
    }

    \label{FIG:P_surv}
\end{figure}

The simulations are based on a basic Monte Carlo approach (generation and geometrical propagation), taking into account only the photon component of the EC de-excitation energy \footnote{The penetration depth of electrons with energies below $3 \, \mathrm{keV}$ is in the order of 10 - 100\,nm and can therefore be neglected \cite{X-ray_data}.}. 
Photons with energies below $3 \, \mathrm{keV}$ undergo photoelectric effect and Compton scattering, the first one being the dominating energy loss mechanism.
For each event, a random value of the de-excitation energy $E_\mathrm{EC}$ is chosen, according to the energy distribution given by the $^{163}$Ho EC spectrum. A photon with such energy is then created in a random point of the two-dimensional implantation layer (with a size of $150 \, \mathrm{\upmu m} \times 150 \, \mathrm{\upmu m}$) and a random direction in the 4$\uppi$ sphere is assigned to it. Finally, the probability $P$ that the photon escapes is calculated as $P = \exp({-\mu_\mathrm{m} \rho x})$, where $\mu_{\mathrm{m}}$ is the mass attenuation coefficient \cite{NIS00}, which depends on the energy $E_\mathrm{EC}$, $\rho$ is the density of gold and $x$ is the distance between the origin point and the surface of the absorber.
Different values of absorber thickness in the range between $0.75 \, \mathrm{\upmu m}$ and $5 \, \mathrm{\upmu m}$ and two different energy ranges, namely between $1.8 \, \mathrm{keV}$ and $Q_{\mathrm{EC}} = 2.833 \, \mathrm{keV}$ and between $2.6 \, \mathrm{keV}$ and $Q_{\mathrm{EC}} = 2.833 \, \mathrm{keV}$, have been taken into consideration. 
The simulation has been run with $10^9$ events for each different value of absorber thickness and for each energy range. The output of these simulations consists of a histogram of the escape probability as a function of energy, for each asborber thickness value and for each energy range, as shown in figure \ref{FIG:P_surv}. The integral of such histogram gives the total escape probability for a given absorber thickness $t$, and the corresponding stopping power $S(t)$ is

\begin{equation}
S(t) = 1 - \int_{E_\mathrm{min}}^{E_\mathrm{max}} P(E,t) \mathrm{d}E
\end{equation}

\noindent where $E_\mathrm{min}$ and $E_\mathrm{max} = Q_{\mathrm{EC}}$ are the energy boundaries in the simulation.

\begin{figure}[h!]
    \centering
    \begin{subfigure}[b]{0.49\textwidth}
        \centering
        \includegraphics[trim=12 0 12 0,clip,width=\linewidth]{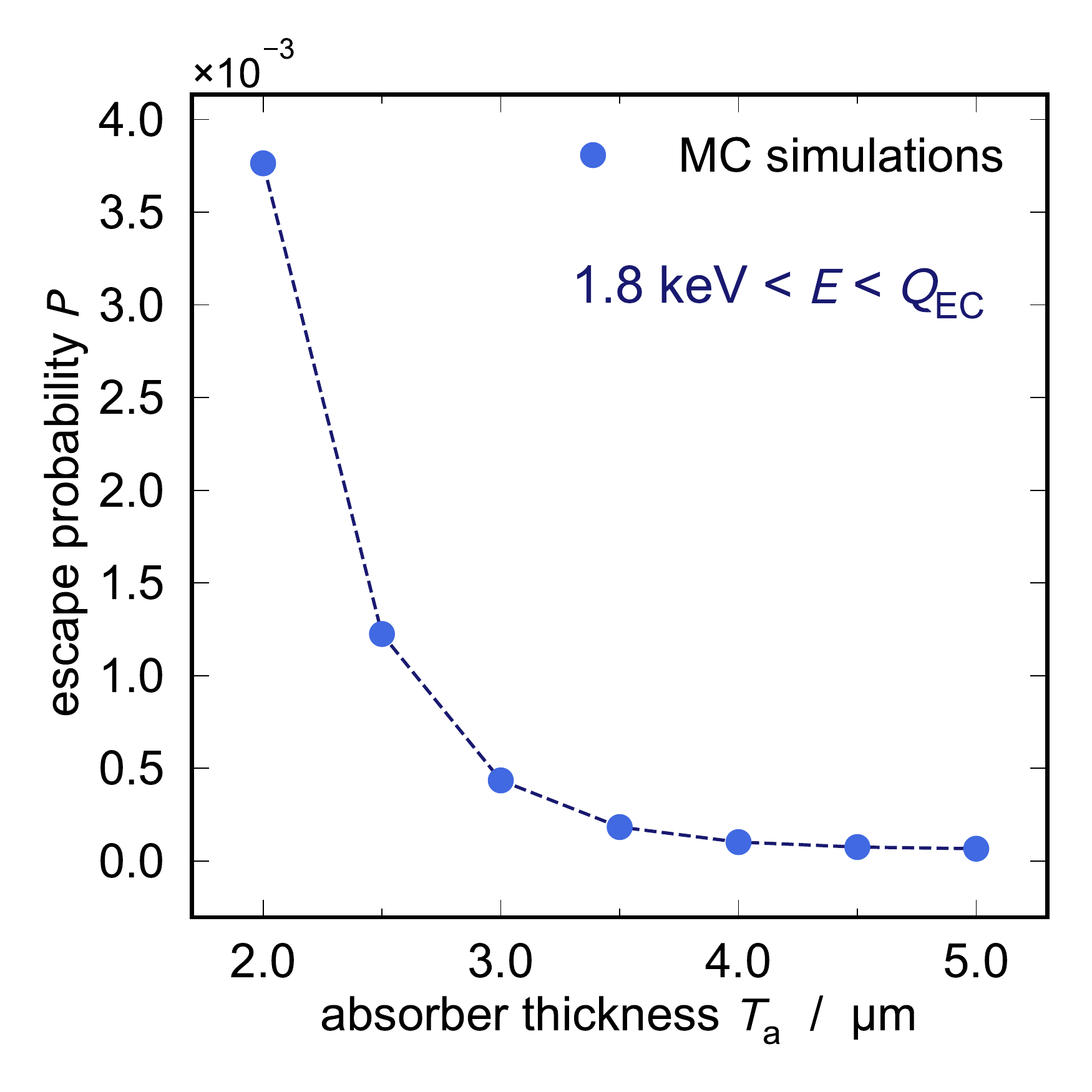}
        \caption{} \label{SUBFIG:surv_prob_vs_thickness}
    \end{subfigure}
    \hfill
    \begin{subfigure}[b]{0.49\textwidth} 
        \centering
        \includegraphics[trim=12 0 12 0,clip,width=\linewidth]{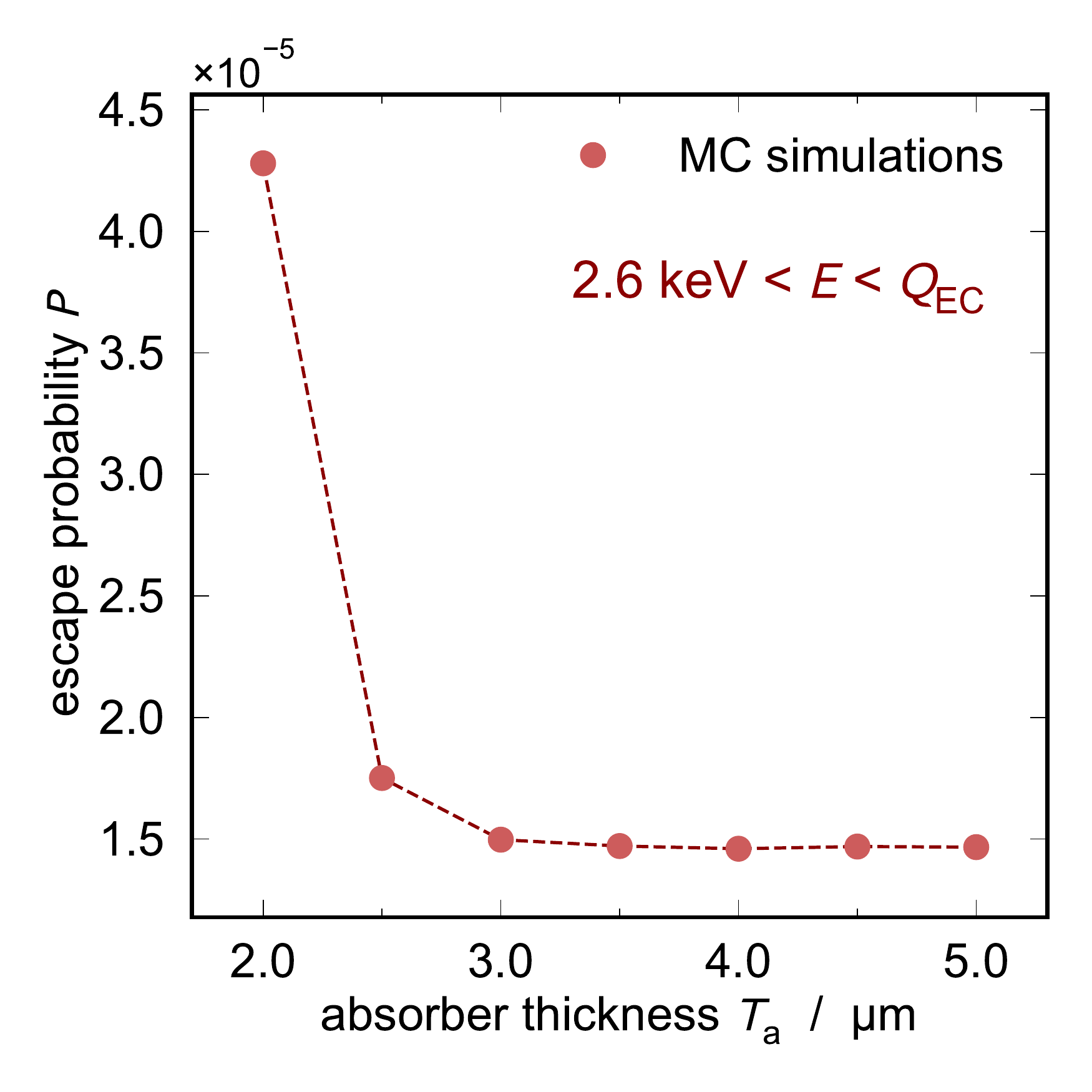}
        \caption{} \label{SUBFIG:surv_prob_endpoint_vs_thickness}
    \end{subfigure}

    \caption{Escape probability resulting from the Monte Carlo simulations as a function of the absorber thickness $T_{\mathrm{a}}$, for the energy range $1.800 \, \mathrm{keV} - Q_{\mathrm{EC}}$ in \textbf{(a)} and for the energy range $2.600 \, \mathrm{keV} - Q_{\mathrm{EC}}$ in \textbf{(b)}.}

    \label{FIG:surv_probability_vs_thickness}
\end{figure}

Figure \ref{FIG:surv_probability_vs_thickness} shows the summary plots of the escape probability as a function of the absorber thickness for the two considered energy ranges.
Since the final goal of the ECHo experiment is the analysis of the end-point region of the $^{163}$Ho EC spectrum to investigate the effective electron neutrino mass, it is crucial to minimise the escape probability at energies above $2.6 \, \mathrm{keV}$. 
On the other hand, in order to test the theoretical approach to the description of the $^{163}$Ho spectrum currently under development \cite{ECHo_spectrum_2019} \cite{Brass_HoSpectrum_2020}, it is important to reliably measure the spectrum also at lower energies. 

\newpage
Overall, an absorber thickness of $2.5 \, \mathrm{\upmu m}$ already guarantees sufficient stopping power, with the escape probability for the events corresponding to the MI line (i.e.~an energy range between $2.02 \, \mathrm{keV}$ and $2.06 \, \mathrm{keV}$) being $9.86 \times 10^{-4}$. 
Assuming a total statistics of $10^{13}$ events, as planned for ECHo-100k, the number of events falling in the MI line region is about $2.2 \times 10^{12}$ and only $2.1 \times 10^{9}$ would be undetected, not significantly affecting the line shape. On the other side, looking at the end of the region of interest for the final analysis, i.e.~the last $50 \, \mathrm{eV}$ at the end-point, the escape probability is estimated to be about $1.65 \times 10^{-8}$. In absolute numbers, for a total statistics of $10^{13}$ events, about $8.3 \times 10^{5}$ are present in this range and no events are expected to be undetected. 
A drawback of a very thin absorber is the potential mechanical instability during the micro\-fa\-bri\-cation processes and during the delicate implantation post-processing. 
In particular, structures with a thickness in the range of $2.0-2.5 \, \mathrm{\upmu m}$ have shown to be prone to collapses or damages, especially during microfabrication steps involving an ultrasonic bath. Thus, a compromise choice is needed in order to minimise the absorber volume, and thus the absorber heat capacity, while maintaining sufficient stopping power and sufficient mechanical stability. 
The final absorber thickness value chosen for the design of the ECHo-100k detector is $3.0 \, \mathrm{\upmu m}$, which on one hand guarantees sufficient stopping power and on the other hand does not endanger the detector stability, as demonstrated with dedicated stability tests \cite{Mantegazzini2021}.

\subsubsection*{Suppression of energy loss}

Quantum efficiency is a crucial detector property for the success of the ECHo experiment. In fact, missing events or missing energy - especially in the end-point region of the $^{163}$Ho spectrum - would compromise the correct reconstruction of the neutrino mass. As discussed in section \ref{SUBSEC:design}, the absorber stands on gold pillars to minimise the contact area to the sensor in order to lower the probability of loss of energy due to athermal phonons travelling through the sensor. However, if an electron capture event occurs in the absorber volume directly above a pillar, athermal phonons could potentially travel to the sensor, which is in direct contact with the pillar's base. In order to avoid this situation, in the ECHo-100k design the layout of the implantation area of each pixel consists of a square with an area of $150 \times 150 \, \upmu \mathrm{m}^2$, with the areas directly above the pillars left not implanted. In this way, the probability of loss of energy due to the mechanism explained above is minimised \cite{Hassel2016}.

\subsubsection*{On-chip thermalisation}

\begin{figure}[h!]
    \centering
    \begin{subfigure}[b]{0.4\textwidth}
        \centering
        \includegraphics[width=0.98\linewidth]{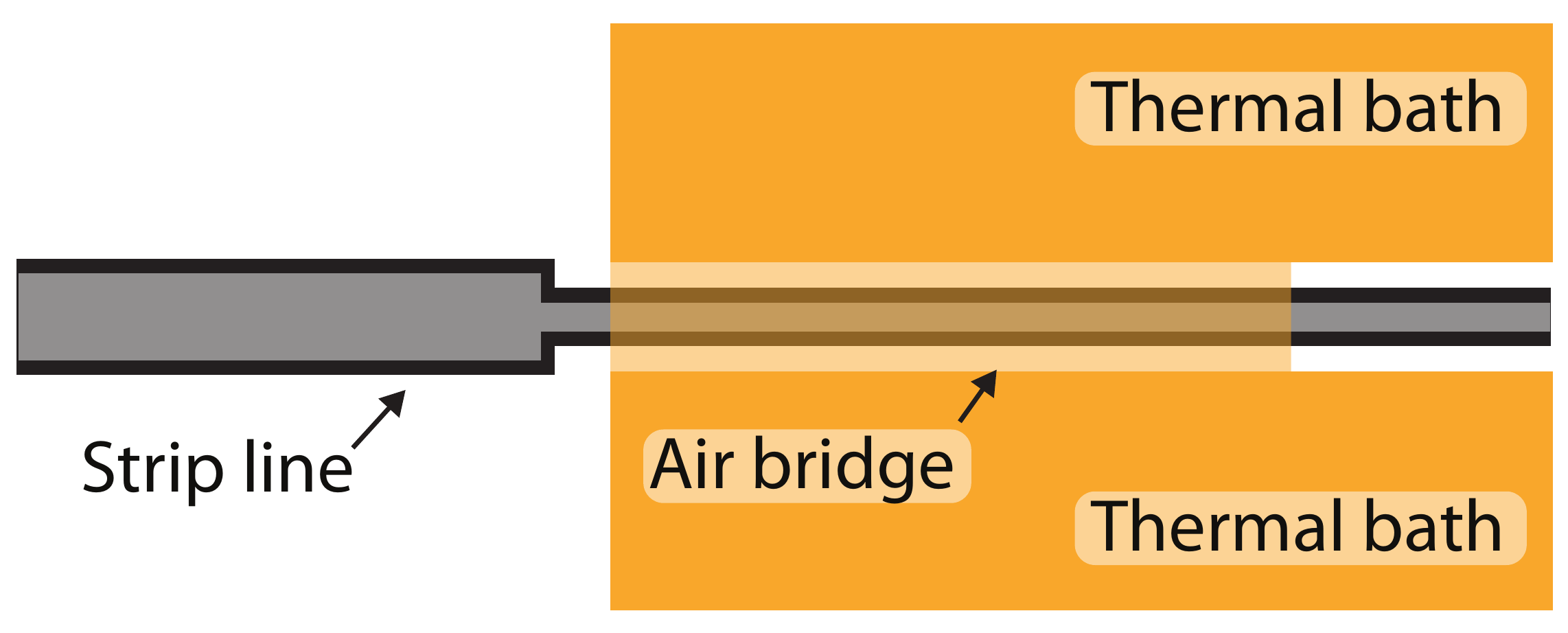}
        \caption{} \label{SUBFIG:air_bridge_top}
    \end{subfigure}
    \hfill
    \begin{subfigure}[b]{0.4\textwidth} 
        \centering
        \includegraphics[width=0.99\linewidth]{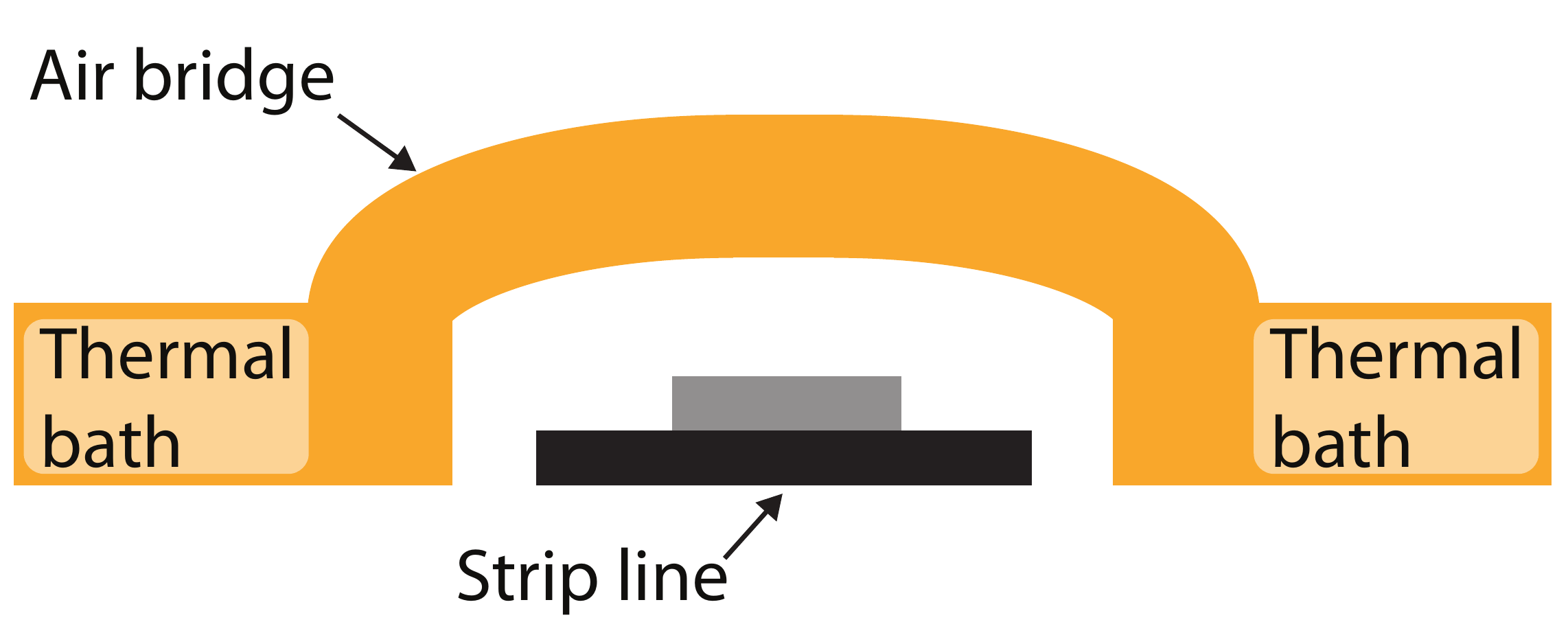}
        \caption{} \label{SUBFIG:air_bridge_side}
    \end{subfigure}
    \caption{Top \textbf{(a)} and side \textbf{(b)} views of the air bridge structure to connect consecutive thermal baths in the ECHo-100k design. The thermal baths are separated by niobium strip-lines (consisting of two overlapping niobium lines with isolation in between).}
    \label{FIG:air_bridge}
\end{figure}

The pulse shape and in particular the signal decay depend on the thermalisation mechanisms that allow the detector to reach the base temperature after an energy deposition event. 
In order to improve the detector thermalisation, the on-chip thermalisation bath features an improved geometry with respect to previous designs \cite{ECHo-1k}. In particular, the gold thermalisation bath areas of the single MMC pixels are connected via gold air bridges that are structured during the microfabrication process, as explained in section \ref{SUBSEC:microfabrication}. A schematic top view and side view of the microfabricated air bridge structure are depicted in figure \ref{FIG:air_bridge}. This innovative solution avoids the usage of potentially unstable and easily detachable gold wire-bonds to connect consecutive thermalisation baths. 

\subsubsection*{Multichip operation}

\begin{figure}[h!]
    \centering
    \begin{subfigure}[b]{0.496\textwidth}
        \centering
        \includegraphics[width=\linewidth]{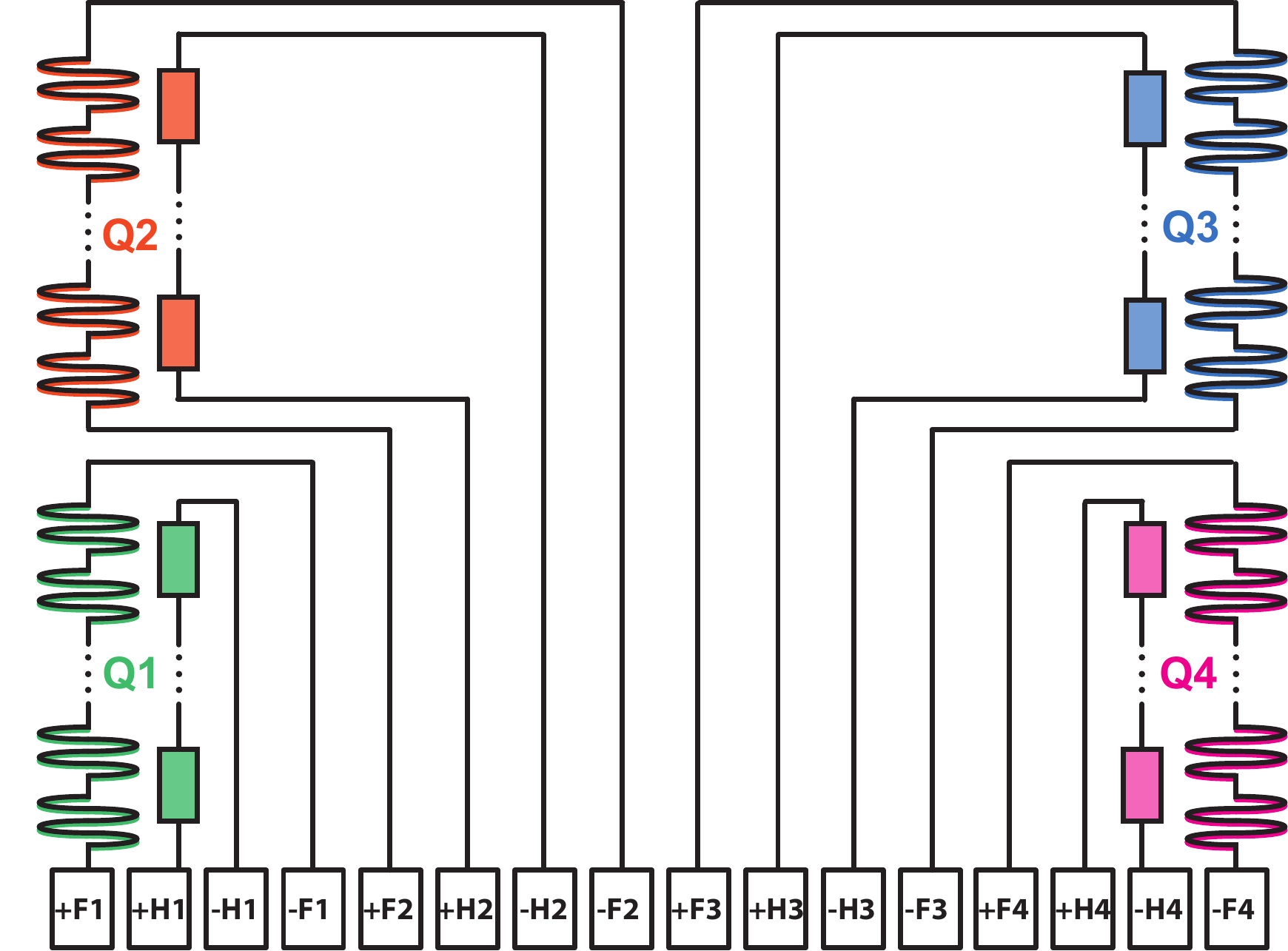}
        \caption{} \label{SUBFIG:ECHo-1k_F_H}
    \end{subfigure}
    \hfill
    \begin{subfigure}[b]{0.496\textwidth} 
        \centering
        \includegraphics[width=\linewidth]{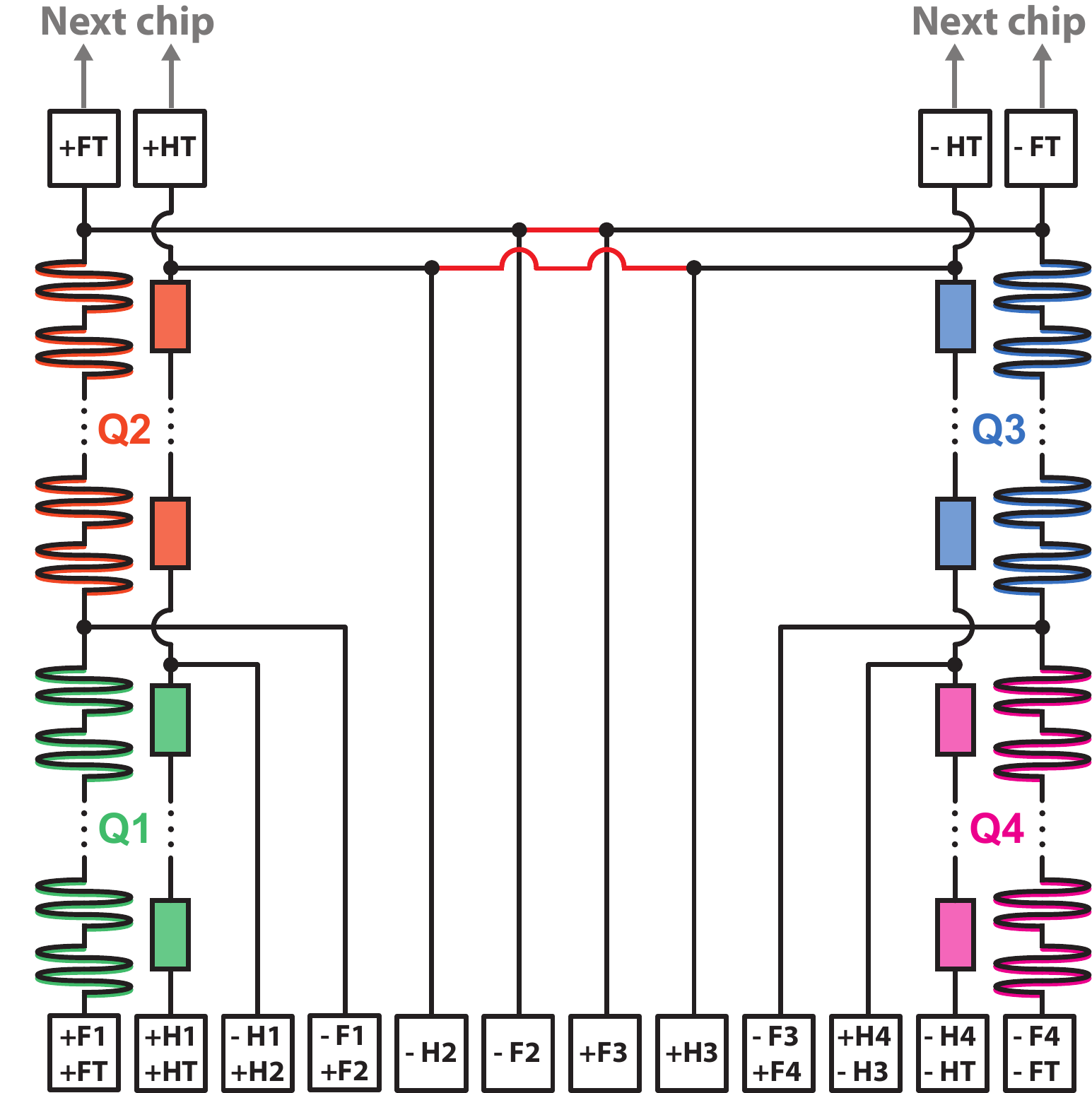}
        \caption{} \label{SUBFIG:ECHo-100k_F_H}
    \end{subfigure}
    \caption{On-chip circuitry dedicated to the injection of the persistent current in the superconducting pick-up coils for the ECHo-1k design \textbf{(a)} and the upgraded ECHo-100k design \textbf{(b)}.}
    \label{FIG:F_H}
\end{figure}

In order to read out thousands of MMC pixels exploiting a multiplexed scheme, the detector design must allow for multichip operation, i.e.~daisy chained detector chips. In particular, it is necessary to inject the persistent current in the superconducting pick-up coils of all the detector chips in the chain at once. 
In the ECHo-1k design the 72 MMC pixels (i.e.~36 read-out channels) are divided in four quarters, each one consisting of 18 MMC pixels (i.e.~9 read-out channels) with the superconducting coils connected in series, as depicted in the schematics of figure \ref{SUBFIG:ECHo-1k_F_H}. Thus, the procedure for the injection of the persistent current must be performed quarter by quarter for each detector chip \cite{ECHo-1k}, using the corresponding bond-pads (e.g.~$\pm \mathrm{F_2}$ and $\pm \mathrm{H_2}$ to inject current into the second quarter).
In the ECHo-100k design the 64 MMC pixels (i.e.~32 read-out channels) are divided in four quarters, each one consisting of 16 MMC pixels (i.e.~8 read-out channels) and the on-chip circuitry has been upgraded in order to allow for the injection of the persistent current with three different approaches, namely 1) injection quarter by quarter in one detector chip, 2) injection into all the pixels of a detector chip at once, 3) injection into all the pixels of multiple detector chips at once.
Figure \ref{SUBFIG:ECHo-100k_F_H} shows the schematics of the on-chip circuitry of the ECHo-100k design.

The first option allows to inject different values of persistent current in different pixels, so that the corresponding detector responses can be efficiently acquired in parallel and compared. In this case, the pads labelled according to the corresponding quarter number (e.g.~$\pm \mathrm{F_2}$ and $\pm \mathrm{H_2}$) are used.
The second option guarantees a homogeneous preparation of the persistent current in the complete detector chip. In this case, the bond-pads $\pm \mathrm{F_T}$ and $\pm \mathrm{H_T}$ are used and the wire-bonds which close the circuit - indicated in red in the scheme of figure \ref{SUBFIG:ECHo-100k_F_H} - need to be placed.
The third option can be implemented if many detector chips are daisy chained in series in a multiplexing set-up, where the persistent current can be homogeneously inserted in all the detector chips simultaneously. Several detector chips can be connected placing aluminium wire-bonds from each chip to the consecutive one, as shown in figure \ref{SUBFIG:ECHo-100k_F_H}. The last chip of the series needs to be equipped with the wire-bonds which close the circuit. In this case, the bond-pads $\pm \mathrm{F_T}$ and $\pm \mathrm{H_T}$ of the first chip of the series are used.

\subsection{Fabrication}
\label{SUBSEC:microfabrication}

In each wafer, 40 ECHo-100k chips are fabricated, organised in four rows. 
At the periphery of the wafer test structures, useful to control the quality of each single layer, are placed.
The fabrication steps and the micro-lithography techniques employed for each layer are equivalent to the ones used for the production of the ECHo-1k detectors \cite{ECHo-1k}.
The complete list of fabrication steps is reported in the appendix, table \ref{TAB:fabrication} and in figure \ref{FIG:ECHo-100k_fabrication_steps}.
Particular care has to be taken for the fabrication of the thermalisation air bridges (layer 8) and of the absorbers (layer 10-14). More details are provided in the appendix, section \ref{SEC:appendix}.

A selection of the fabricated ECHo-100k chips with only the first absorber layer deposited (layer 10) have been characterised before proceeding with the $^{163}$Ho implantation and the deposition of the second absorber layer (layers 11-14).

\section{Characterisation} \label{SEC:characterisation}


\subsection{Characterisation at room temperature}

After fabrication, the complete ECHo-100k wafers underwent a detailed optical inspection with the microscope to check the continuity of the niobium lines as well as the integrity of the main structures. 
As next step, the resistances of the niobium lines have been tested at room temperature, manually contacting the corresponding bond-pads with a needle-prober tool\footnote{SUSS PA200 Semiautomatic Prober System, Süss MicoTec}.
The read-out connections feature resistances in the range between $915 \, \mathrm{\Omega}$ and $1060 \, \mathrm{\Omega}$, depending on the length of the niobium lines.
The connections for the injection of persistent currents feature larger resistances in the range between $2.87 \, \mathrm{k\Omega}$ and $4.41 \, \mathrm{k\Omega}$, due to the long paths through the meander-shaped pick-up coils.
In order to make the needle-prober measurement more efficient for higher numbers of wafers, a dedicated automated script has been developed, which allows to probe all the lines of all of the chips on the wafer in a fully automatic fashion.

\subsection{Characterisation at 4 \, K}
\label{4K_characterisation}

The detector chips have been cooled down to about $4 \, \mathrm{K}$ by means of a liquid helium bath or of a pulse tube cryocooler, in order to measure resistances, critical current, heater switches and inductance of the pick-up coil. 

\paragraph{Resistances}

To prove the superconductivity of the read-out niobium lines and of the lines connecting the detector meander-shaped pick-up coils, as well as to measure the resis\-tances of the gold-palladium heaters, four-wire measurements were performed at a temperature of $4.2 \, \mathrm{K}$. From these measurements it is possible to estimate the resistance of a single heater element, $R_\mathrm{H}(T = 4.2 \, \mathrm{K}) = 6.3 \pm 0.4 \, \mathrm{Ohm}$, which is compatible with the expectations based on the material and the geometry of the heaters \cite{Hengstler_dissertation}.

\paragraph{Critical current}

The critical current of the niobium structures that form the meander-shaped pick-up coils can be measured running a current through the connections that link all the pick-up coils belonging to one quarter of the chip and monitoring the corresponding resistance via a four-wire measurement.
The average critical current is estimated to be $\bar{I_\mathrm{c}}(T = 4.2 \, \mathrm{K}) = 102.8 \pm 4.4 \, \mathrm{mA}$ in a liquid helium bath.
The typical persistent current values that are used to operate the detector are below $50 \, \mathrm{mA}$ and therefore about a factor of two smaller than the measured critical current.
With the same approach, it is possible to determine the critical current through the vias\footnote{The term "via" refers to a vertical electrical connection between different layers. In this particular case, the vias are connecting the niobium layers "SQUID lines (layer 1)" and "SQUID lines (layer 2)", as reported in table \ref{TAB:fabrication}}, which feature a size of $12 \, \mathrm{\upmu m} \times 12 \, \mathrm{\upmu m}$. 
The average critical current through vias is $\bar{I_\mathrm{c}}_\mathrm{,vias}(T = 4.2 \, \mathrm{K}) = 38 \pm 1 \, \mathrm{mA}$, which is sufficient, as vias are only present in the connection lines routed to the input coil of the SQUID, where the maximum flowing current is in the order of tens of micro-ampere.

\paragraph{Heater switch}

The area of the heater switch in the ECHo-100k design ($20 \, \mathrm{\upmu m} \times 5 \, \mathrm{\upmu m}$) is reduced with respect to the ECHo-1k design ($30 \, \mathrm{\upmu m} \times 5 \, \mathrm{\upmu m}$) and its position is closer to the niobium circuit. 
The minimum current value that activates the heater switch is therefore expected to be different from the case of the ECHo-1k detector.
This parameter can be determined via a four-wire measurement of the heater circuit by running different currents through it.
The average minimum current that activates the heater switch has been measured in a liquid helium bath and the result is $\bar{I}_\mathrm{h,min} \approx \mathrm{2.0 \pm 0.3 \, mA}$.

\paragraph{Inductance of the pick-up coil}

The inductance of the niobium superconducting meander-shaped pick-up coils is an essential parameter for the flux coupling between detector and SQUID.
Considering the circuit formed by the meander-shaped pick-up coil, the aluminium wire-bonds and the input coil of the SQUID, a flux noise measurement of the SQUID performed at $T \approx \mathrm{4 \, K}$ can be used to determine the inductance of the pick-up coil \cite{ECHo-1k}.
The methodology, the experimental data and the corresponding fit are discussed in the appendix, section \ref{SEC:appendix}.
The obtained value for the pick-up coil inductance is $L_\mathrm{m} = 2.1 \pm 0.2 \, \mathrm{nH}$ and is consistent with the expected value of $L_\mathrm{m, sim} = 2.27 \, \mathrm{nH}$, obtained from simulations\footnote{The pick-up coil inductance has been simulated with InductEx, \url{http://www0.sun.ac.za/ix}.}.

\subsection{Characterisation at millikelvin}

The cryogenic set-up and read-out chain employed during the measurements at millikelvin temperatures have been discussed in \cite{readOut_paper}.

\subsubsection{Magnetisation response}
\label{magnetisation_section}

The measurement of the magnetisation of the paramagnetic sensor follows the same procedure as discussed in \cite{ECHo-1k}, by changing the temperature of the mixing chamber of the cryostat in a range between $\mathrm{10 \, mK}$ and $\mathrm{400 \, mK}$, and acquiring the SQUID voltage output of the non-gradiometric channels. \\
The measured magnetisation values are converted from units of V to the units of flux quantum $\Phi_0$ using the experimentally determined voltage-to-flux conversion factor $V/\Phi_0$, which was determined to be $0.377 \, \mathrm{V/\Phi_0}$\footnote{Voltage-to-flux conversion factor can differ from one SQUID to another due to slight differences in mutual inductance, and therefore needs to be determined for each SQUID separately.}. More detailed description of the magnetisation measurement is provided in the appendix, section \ref{SEC:appendix}. \\
Figure \ref{fig:magnetisation_3} shows the magnetisation curves obtained for the ECHo-100k detector. 
The measurements have been performed for three different values of the persistent current in the meander-shaped pick-up coils, namely $\mathrm{20 \, mA}$, $\mathrm{40 \, mA}$ and $\mathrm{50 \, mA}$. \\

\begin{figure}[h!]
	\centering
    \includegraphics[width=1.0\textwidth]{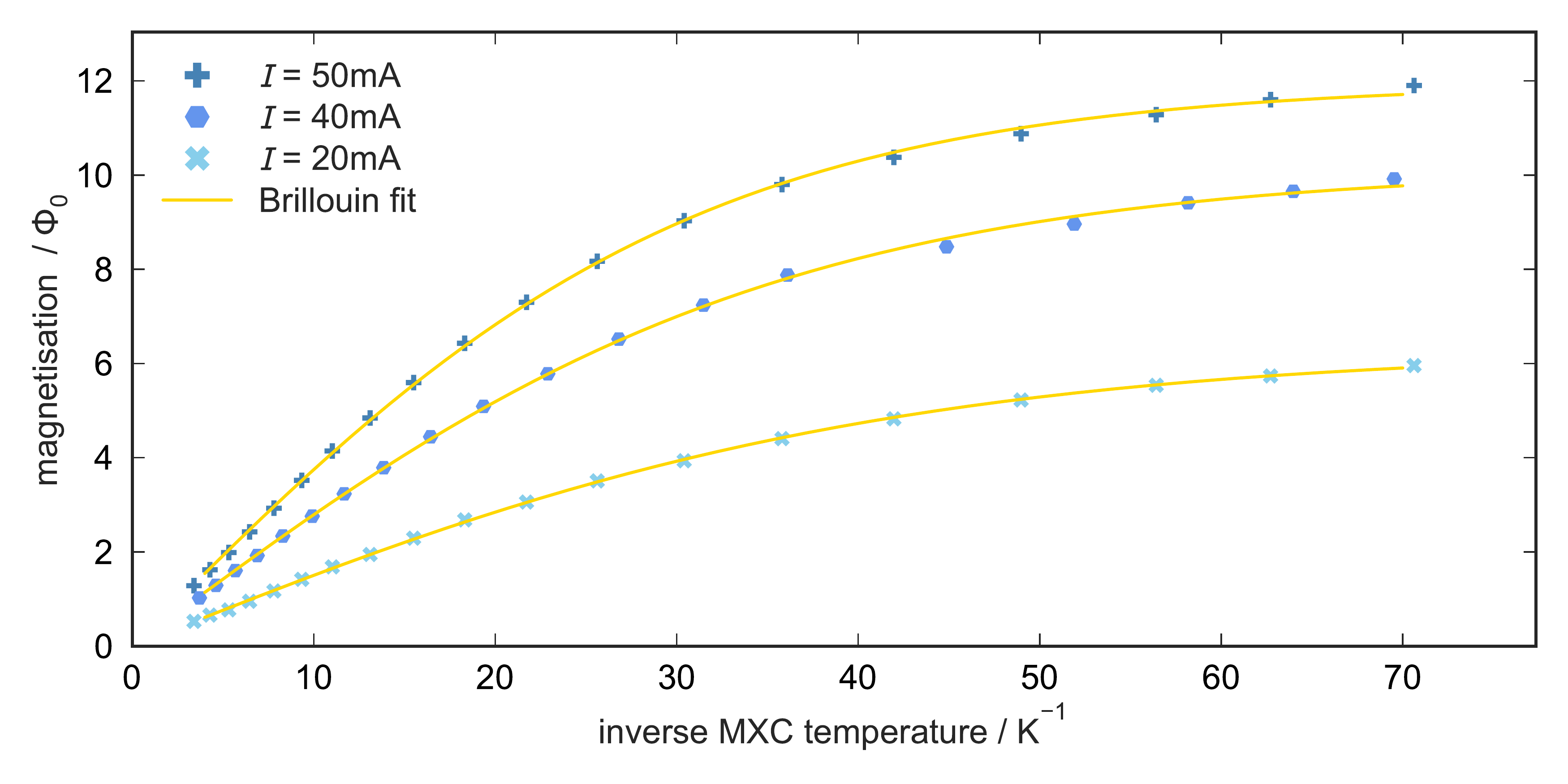} 
	\caption{Magnetisation curves from the non-gradiometric channel of the ECHo-100k detector, in units of flux against inverse temperature, for three different values of the persistent current (20 mA, 40 mA and 50 mA). All curves are fit with the analytic function \ref{equation:magnetisation_expression}, shown in different shades of yellow.}
	\label{fig:magnetisation_3}
\end{figure}

The magnetisation of the spin system of the sensor is given by equation \cite{Fle2005}: 
\begin{equation}
	M \; = \; \frac{N}{V} \tilde{g} \tilde{S} \mu_B \mathcal{B}_{\tilde{S}} \Big( \frac{\tilde{g} \tilde{S} \mu_{\mathrm{B}} {\it B}}{k_{\mathrm{B}} T}\Big) \; = \; k \cdot \mathcal{B}_{\tilde{S}} (aT^{-1}),
	\label{equation:magnetisation_expression}
\end{equation}

\noindent which is used to fit the data, with $k$ and $a$ left as the free parameters.
\noindent $\mathcal{B}_{\tilde{S}} (aT^{-1})$ is the Brillouin function. From the obtained value of the fit parameter $a \, = \, \tilde{g} \tilde{S} \mu_{\mathrm{B}} {\it B} k_{\mathrm{B}}^{-1}$ and the known values of $\tilde{g}_{\mathrm{Ag:Er}}$ = 34/5 \cite{Fle2005}, $\tilde{S}$ = 1/2 \cite{Fle2005}, $\mu_{\mathrm{B}} \,= \, 9.274 \times 10^{-24} \, \mathrm{JT^{-1}}$ and $k_{\mathrm{B}} \, = \, 1.38 \times 10^{-23} \, \mathrm{JK^{-1}} $, the magnetic field $B$ at the position of the sensor can be extracted. The obtained values depend on the persistent current injected in the meander-shaped pick-up coils and on the geometry of the detector.
The values of the average field $B$ obtained from the fit are in the order of $10 \, \mathrm{mT}$, for all values of the persistent current. These values are on the same order of magnitude as the expected values of the magnetic field, obtained from the dedicated simulations performed with the Finite Element Method Magnetics (FEMM) software\footnote{D. C. Meeker, Finite Element Method Magnetics, Version 4.2 (28Feb2018 Build)}. \\ 

\subsubsection{Pulse shape analysis}
\label{pulseShape_section}

The results presented here have been obtained by performing measurements with the non-implanted pixels of the ECHo-100k detectors. 

\begin{figure}[h!]
	\centering
    \includegraphics[width=1.0\textwidth]{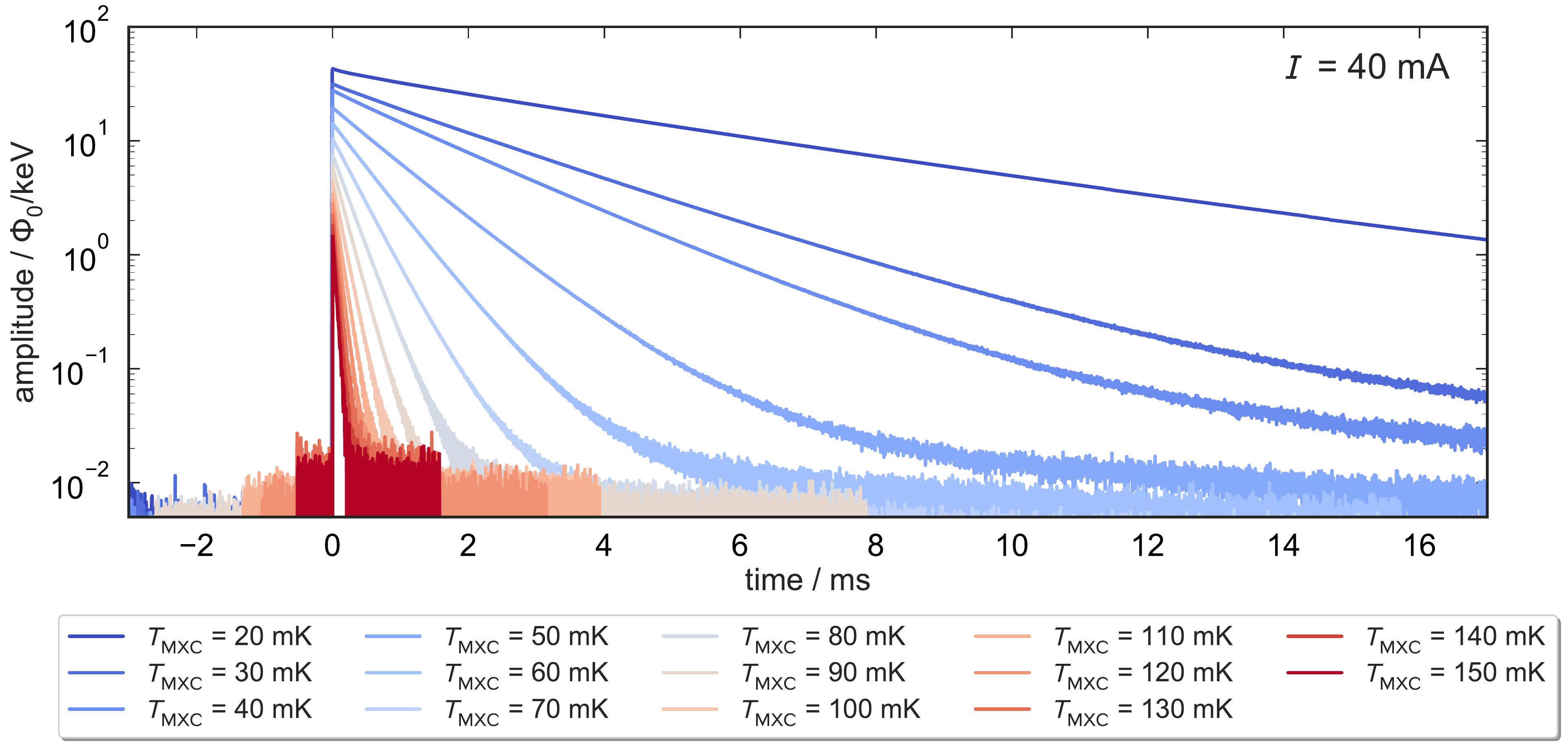} 
	\caption{Pulse shape corresponding to the interaction of $5.89 \, \mathrm{keV}$ photons in an ECHo-100k pixel, for operating temperatures between $20 \, \mathrm{mK}$ and $150 \, \mathrm{mK}$, measured with the pesistent current of 40 mA. The pulses given in units of flux quanta and scaled to the energy of the K$_{\alpha}$ line, and are therefore given in units of $\Phi_0/\mathrm{keV}$. The y-axis is plotted in a logarithmic scale.}
	\label{fig:longTW_temperatureScan}
\end{figure}

In order to reconstruct the detector response, an external $^{55}\mathrm{Fe}$ calibration source illuminating non-implanted pixels has been used.
The corresponding events have been acquired at different temperatures, ranging from $20 \, \mathrm{mK}$ to $150 \, \mathrm{mK}$.
Figure \ref{fig:longTW_temperatureScan} shows how the detector signals corresponding to the $5.89 \, \mathrm{keV}$ K$_{\alpha}$ line change with temperature. The pulse amplitude decreases with the increasing temperature due to the increase of the detector heat capacity, while the decay time of the pulses becomes faster due to the enhanced thermal conductance of the thermal link connecting the pixel to the thermal bath. \\

\paragraph{Pulse amplitudes}

\begin{figure}[h!]
     	\centering
     	\begin{subfigure}[b]{0.49\textwidth}
        	 \centering
         	\includegraphics[width=\textwidth]{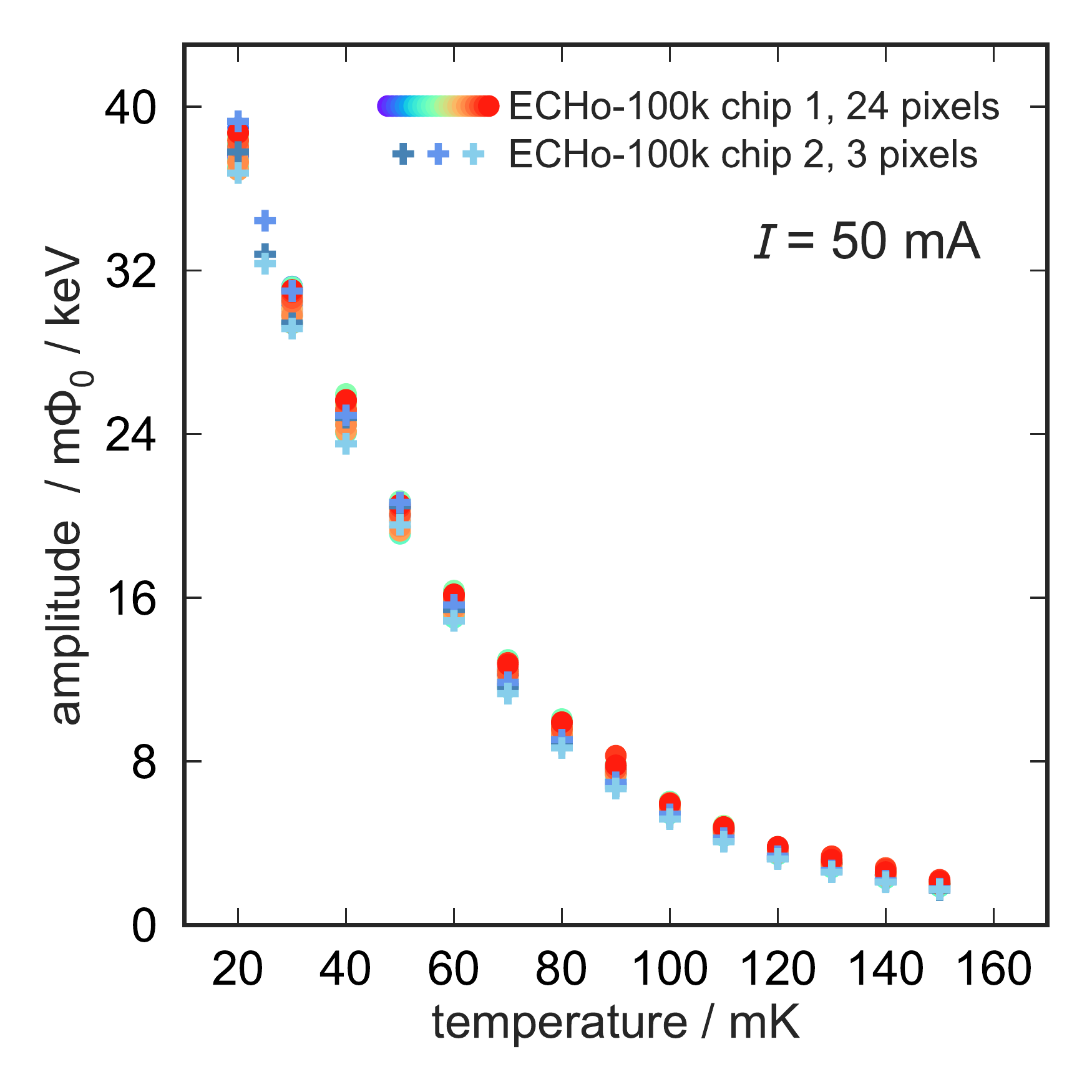}
         	\caption{}
         	\label{subfig:amplitudeComparison}
     	\end{subfigure}
     	\begin{subfigure}[b]{0.49\textwidth}
        	 \centering
         	\includegraphics[width=\textwidth]{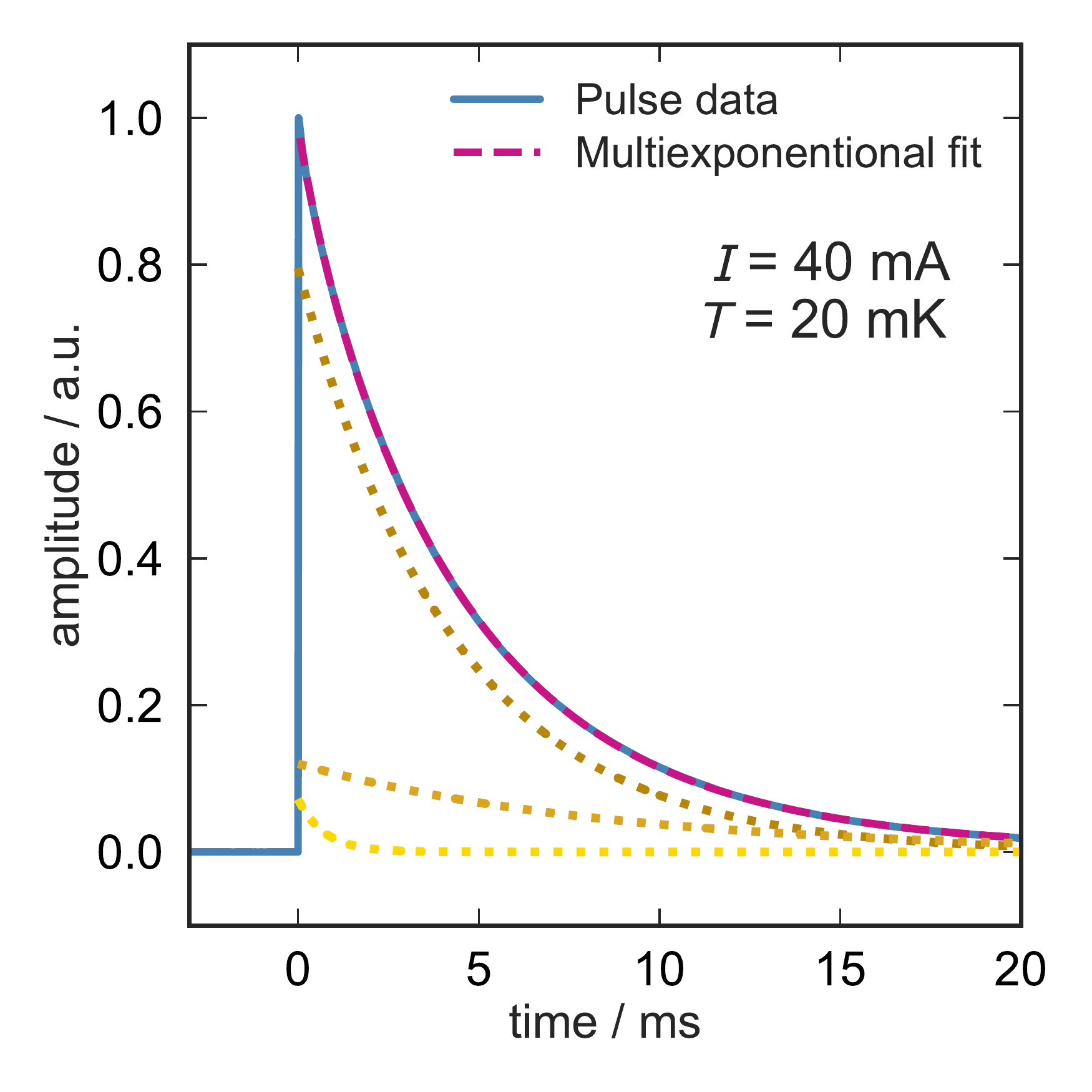}
         	\caption{}
         	\label{subfig:pulseFit}
     	\end{subfigure}
        	
\caption{{\bf (a)} Comparison of the extracted pulse amplitudes, in the temperature range between $20 \, \mathrm{mK}$ and $150 \, \mathrm{mK}$ for two different ECHo-100k chips. {\bf (b)} Example of a multi-exponential fit of a long time window pulse. The individual decay components are plotted in dashed lines. The pulse is scaled to unitary amplitude in order to directly obtain the percentage of the different decay components.}
	\label{fig:amplitude_Comparison}
\end{figure}

Ensuring that the achieved detector performance is reproducible in all fabricated ECHo detectors is of crucial importance for the success of the experiment. \\
Figure \ref{subfig:amplitudeComparison} shows the comparison of the measured amplitudes from the non-implanted pixels of two different ECHo-100k chips. One of the chips did not undergo the implantation procedure and 24 pixels have been measured. A second chip from the same wafer was implanted with the $^{163}$Ho source and therefore only few non-implanted pixels are present on the chip, three of which have been characterised. The amplitudes shown in figure \ref{subfig:amplitudeComparison} correspond to the interaction of $5.89 \, \mathrm{keV}$ photons and are given in units of $\Phi_0 / \mathrm{keV}$. The persistent current in the meander-shaped pick-up coils was $50 \, \mathrm{mA}$ in both measurements. 

The recovery of the initial temperature of the detector is a complex process which can be described by a multi-exponential function, with the largest contribution coming from the thermalisation with the heat sink, and several smaller contributions\footnote{In the case of ECHo-100k detector, a multi-exponential function with three contributions has been shown to successfully describe the signal decay.} attributed to diffusion processes.
In order to extract the amplitude of the signal, the pulse decay is fit with a multi-exponential function, as shown in figure \ref{subfig:pulseFit}, given by $f(t) \; = \; \sum_i  \, A_i \, \exp(-t/\tau_i)$.

The spread of pulse amplitudes from the individual pixels of the same chip ranges from about 6.7\% at $20 \, \mathrm{mK}$ to about 14.8\% at $100 \, \mathrm{mK}$ and can be attributed to the variation of the flux transformer coupling, which depends on the wire-bond inductance as well as on the parasitic inductance, both channel dependent. Additionally, small discrepancies in the geometry of the micro-fabricated structures (e.g.~sensor and absorber volumes) can lead to a slight change in the detector heat capacity, and thus in the signal height. 
Overall, the agreement between the results of the two measurements is excellent as can be seen in the plot of figure \ref{subfig:amplitudeComparison}, with the difference between the mean values of the pulse amplitude of the two measurements being only 0.03\% at 20 mK and going up to 6\% at 100 mK. This serves as a proof of reproducibility of the desired performance and stability of the detector production.

\begin{figure}[t!]
     	\centering
     	\begin{subfigure}[b]{0.49\textwidth}
        	 \centering
         	\includegraphics[width=\textwidth]{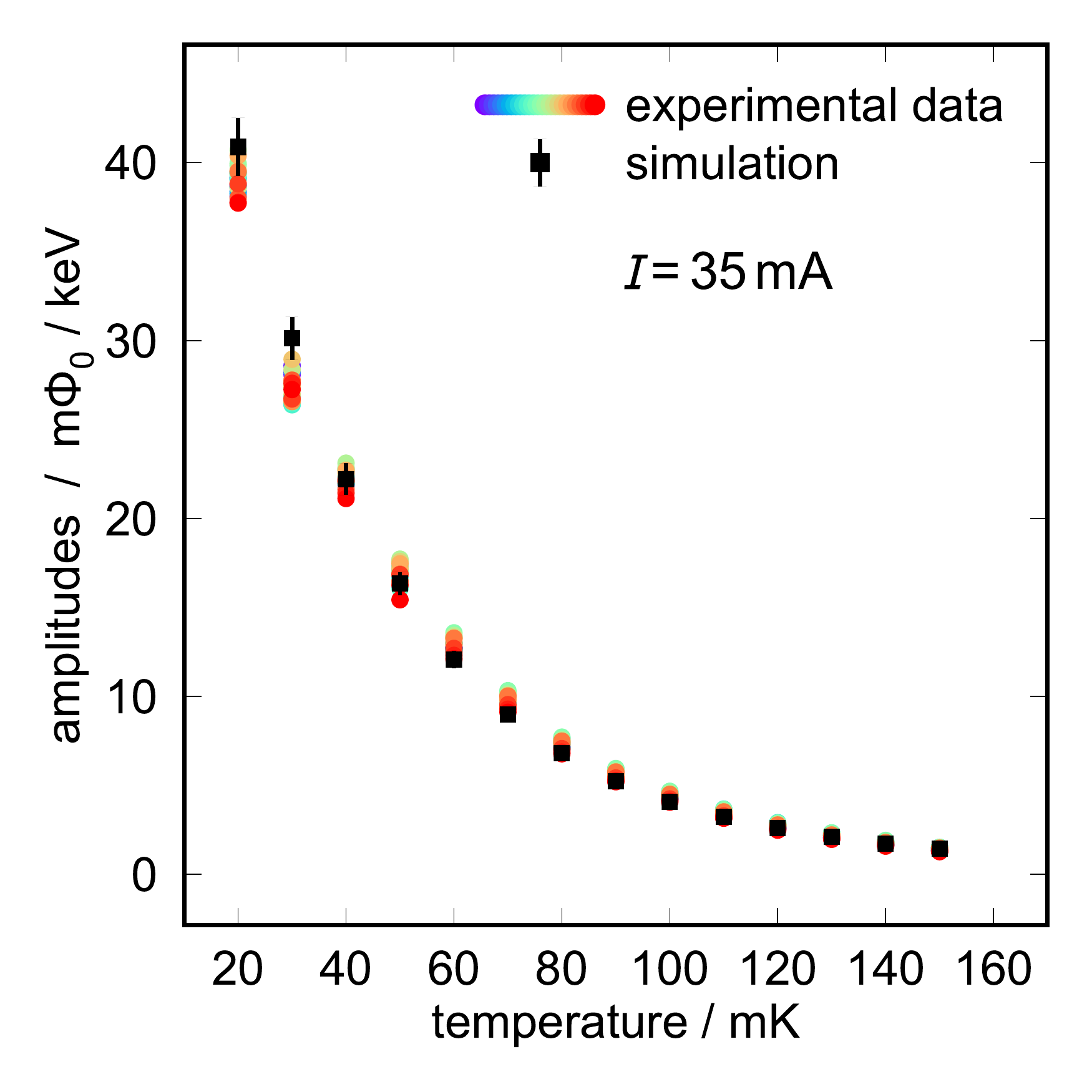}
         	\caption{}
         	\label{theoreticalAmplitudes_35mA}
     	\end{subfigure}
     	\begin{subfigure}[b]{0.49\textwidth}
        	 \centering
         	\includegraphics[width=\textwidth]{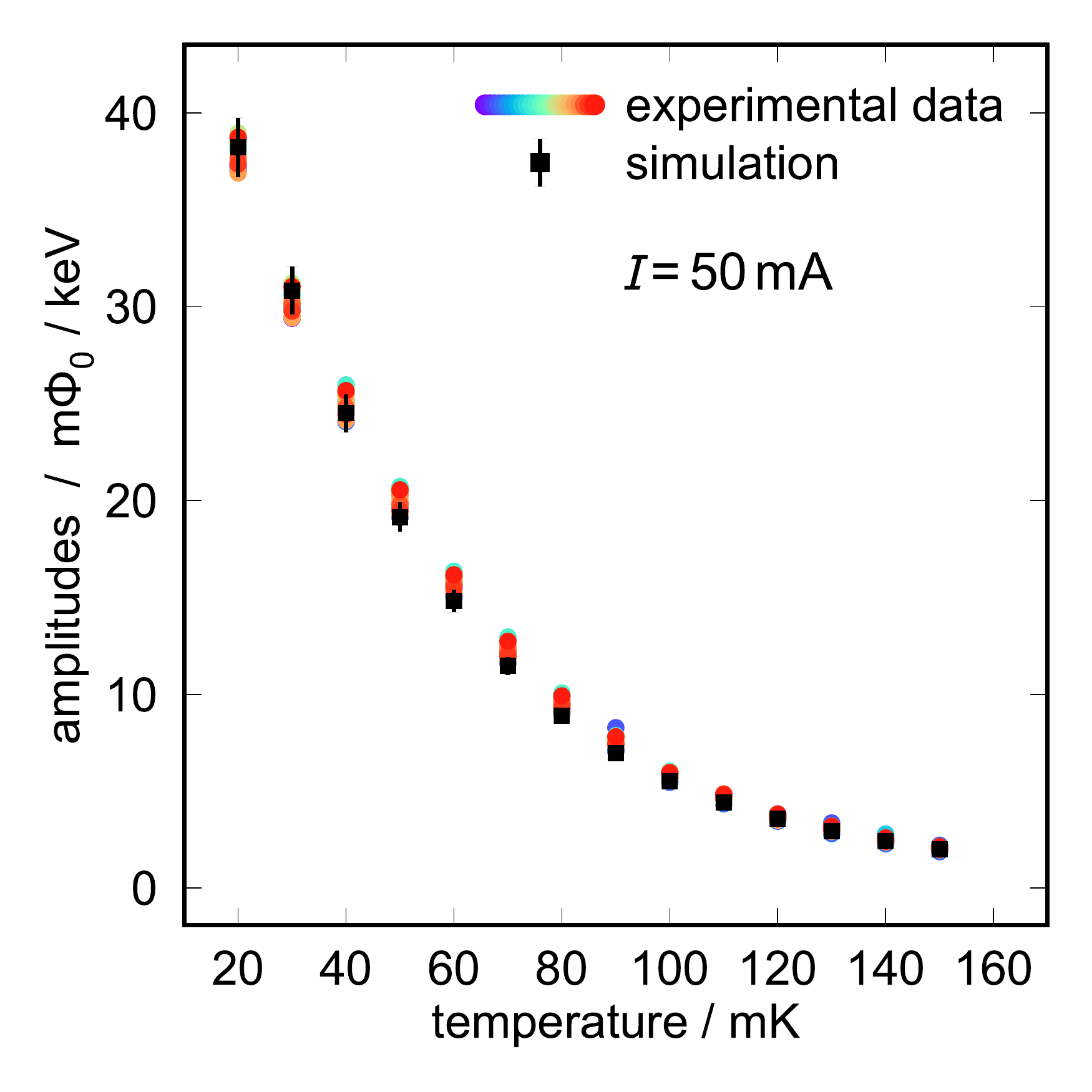}
         	\caption{}
         	\label{theoreticalAmplitudes_50mA}
     	\end{subfigure}
        	
	\caption{The experimentally measured signal amplitudes measured with 24 pixels of the ECHo-100k detector shown in terms of flux in the SQUID per unit of energy (coloured marks), compared with the theoretical expectation based on numerical simulations \cite{Herbst_simulations} (black marks) for a persistent current of $35 \, \mathrm{mA}$ {\bf (a)} and $50 \, \mathrm{mA}$ {\bf (b)}.}
	\label{fig:theoreticalAmplitudes}
\end{figure}

The theoretical amplitudes, calculated from numerical simulations \cite{Herbst_simulations}, are compared with the experimental results in figure \ref{fig:theoreticalAmplitudes} for two different values of the persistent current, namely $35 \, \mathrm{mA}$ and $50 \, \mathrm{mA}$. The amplitudes are larger in the first case, as expected.
The high congruence between experimental data and theoretical expectations demonstrates that the detector response to external energy inputs is well understood.

\paragraph{Detector response to different energy inputs}

\begin{figure}[t!]
     	\centering
     	\begin{subfigure}[b]{0.32\textwidth}
        	 	\centering
         	\includegraphics[height=5cm]{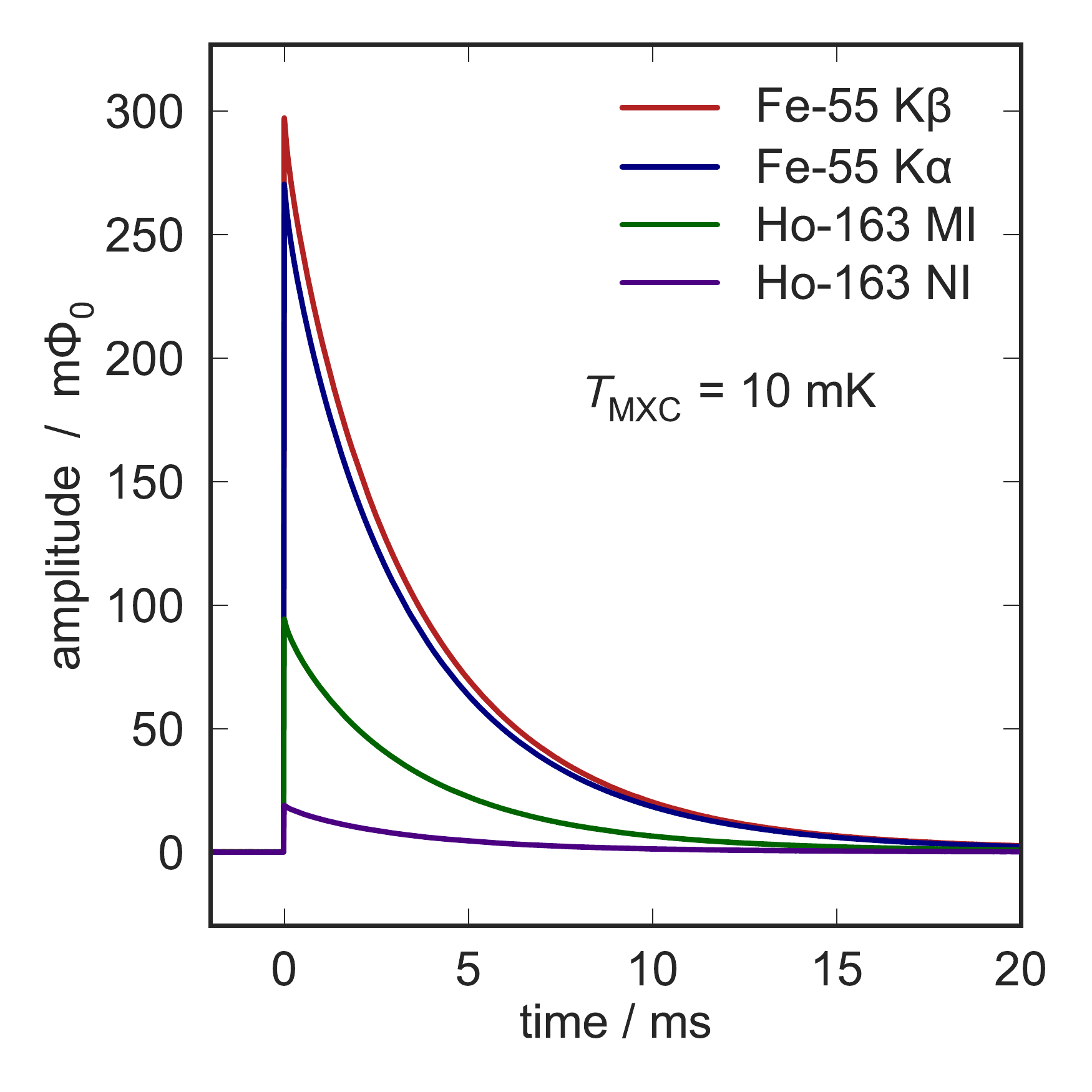}
         	\caption{}
         	\label{subfig:amplitudeComparison_differentLines}
     	\end{subfigure}
     	\begin{subfigure}[b]{0.325\textwidth}
        	 	\centering
         	\includegraphics[height=5cm]{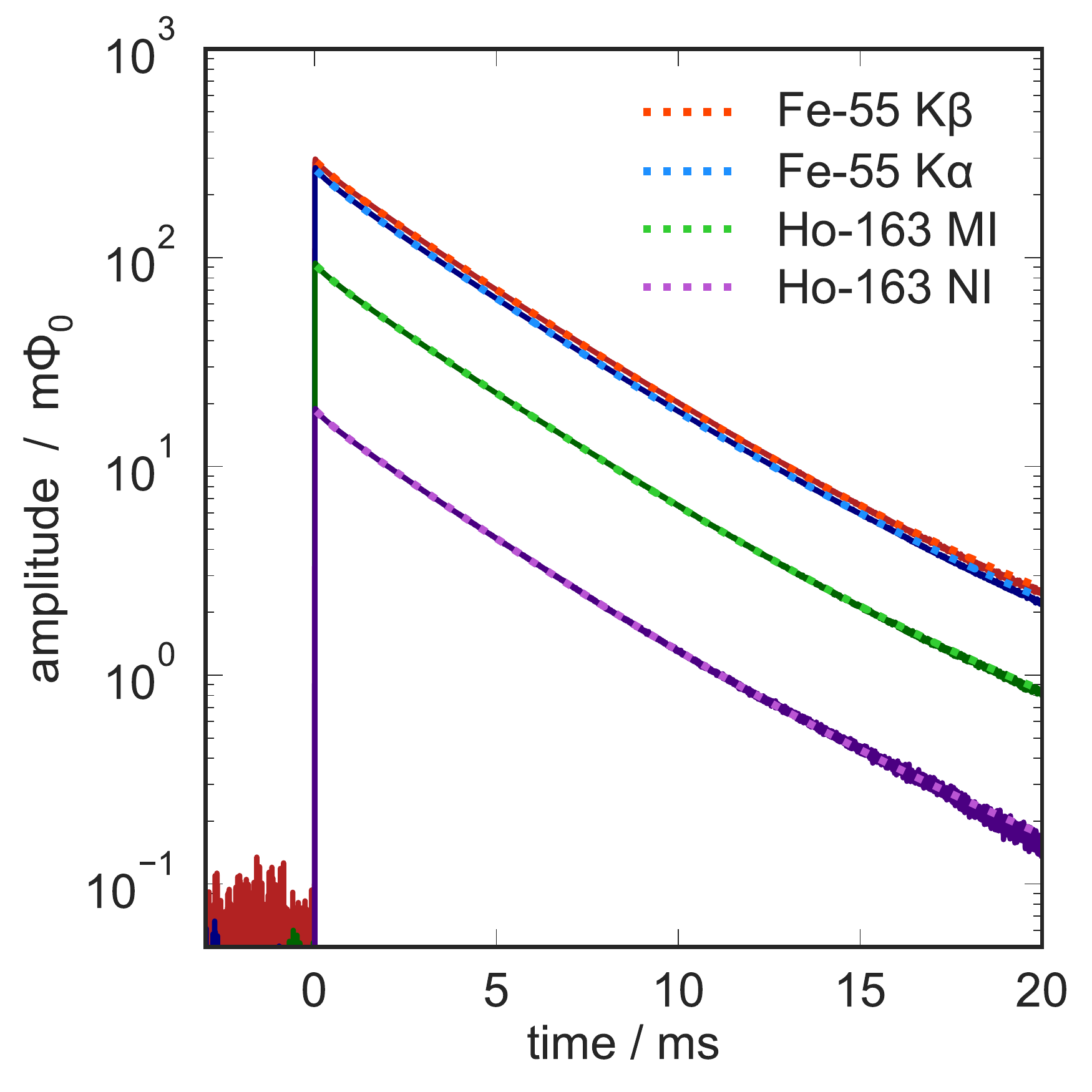}
         	\caption{}
         	\label{subfig:fittedLines_differentLines}
     	\end{subfigure}
     	\begin{subfigure}[b]{0.32\textwidth}
         	\centering
         	\includegraphics[height=5cm]{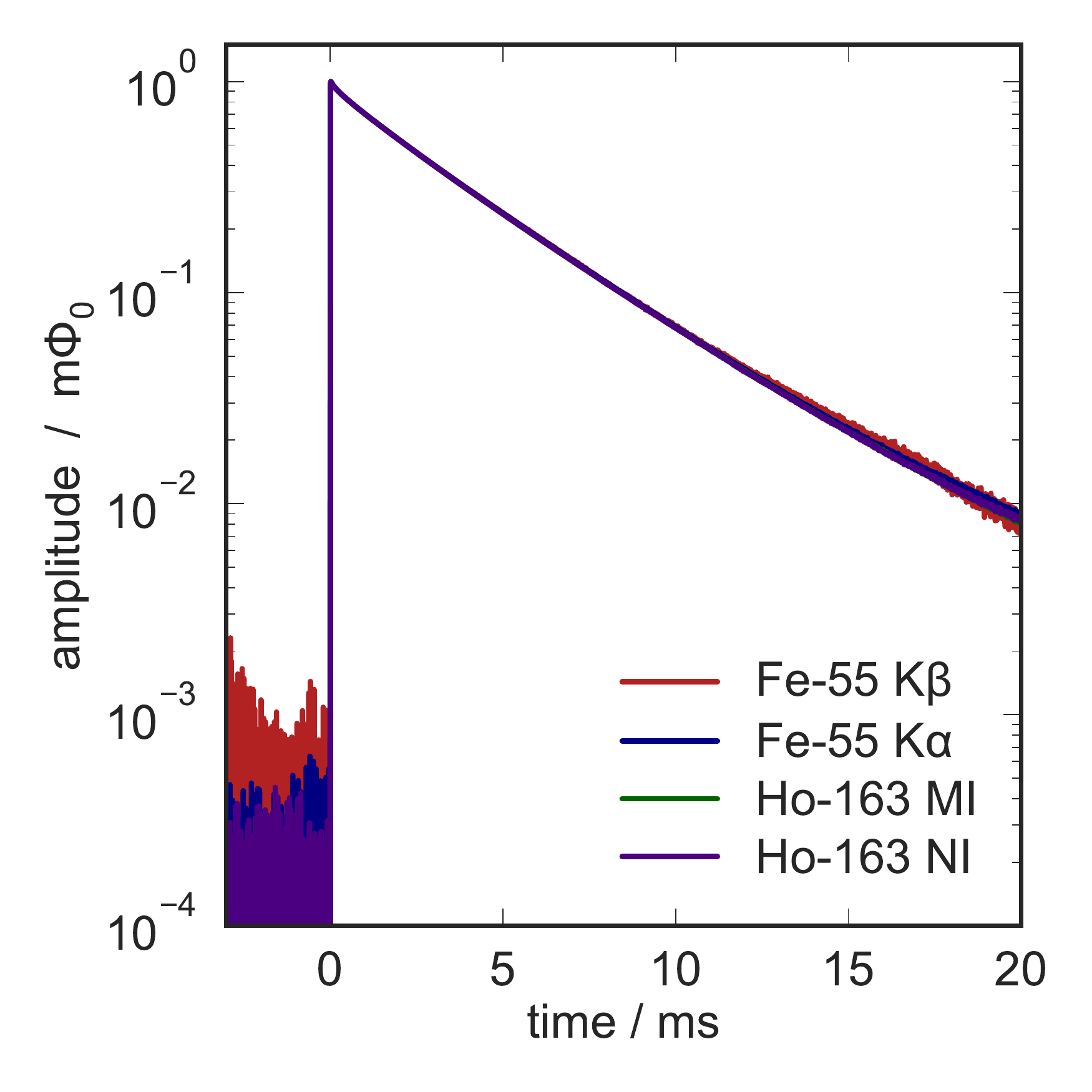}
         	\caption{}
         	\label{subfig:overlayed_differentLines}
     	\end{subfigure}
        	
	\caption{Comparison between measured signals with different energy inputs, namely for the $\mathrm{K}_{\beta}$ and $\mathrm{K}_{\alpha}$ lines of $^{55}\mathrm{Fe}$ and $\mathrm{M}_{\mathrm{I}}$ and $\mathrm{N}_{\mathrm{I}}$ lines of $^{163}\mathrm{Ho}$. Figure {\bf (a)} shows the pulses given in units if $\mathrm{m\Phi_0}$, given in linear scale. Figure {\bf (b)} shows the pulses in logarithmic scale together with the performed multi-exponential fits. Pulses scaled to 1 and overlaid on top of each other are depicted in figure {\bf (c)}.}
    \label{fig:different_lines_comparison}
\end{figure}

In order to prove that the detector response to different energy inputs is consistent, signals from different spectral lines have been selected and compared. As discussed in the previous paragraph, each signal pulse is fit with a multi-exponential function. The results are displayed in figure \ref{fig:different_lines_comparison} for signals measured at about $20 \, \mathrm{mK}$, corresponding to the $\mathrm{K}_{\beta}$ line of $^{55}\mathrm{Fe}$ calibration source, with an energy of $6.49 \, \mathrm{keV}$ \cite{K_alpha_Hoelzer1997}, the $\mathrm{K}_{\alpha}$ line of $^{55}\mathrm{Fe}$ calibration source, with an energy of $5.89 \, \mathrm{keV}$ \cite{K_alpha_Hoelzer1997}, the $\mathrm{MI}$ electron capture line of $^{163}\mathrm{Ho}$ source, with an energy of $2.04 \, \mathrm{keV}$ and, finally, the $\mathrm{NI}$ electron capture line of $^{163}\mathrm{Ho}$ source, with an energy of $0.41 \, \mathrm{keV}$.

Figure \ref{subfig:amplitudeComparison_differentLines} shows that the pulse shape is consistent for different measured lines, as anticipated. Figure \ref{subfig:fittedLines_differentLines} shows these pulses in a logarithmic scale together with the fit performed with a multi-exponential function.
In order to demonstrate the consistency of detector response to different energy inputs, each of the 4 signal pulses shown in figure  \ref{subfig:fittedLines_differentLines} are scaled to 1 and overlaid on top of each other, as depicted in figure \ref{subfig:overlayed_differentLines}, showing a very good agreement between individual signals. \\

\subsubsection{Heat capacity}
\label{heatCapacity_section}

The optimisation process towards the new ECHo-100k design was based on theoretical calculations which assumed a reliable prediction of the single heat capacity contributions, i.e.~heat capacity of absorber, sensor and implanted holmium.
In order to verify the model, the assumed theoretical heat capacity must be compared with the experimentally measured one. 
The heat capacity $C$ of the detector is experimentally accessible from the signal amplitude in temperature $\Delta T \,= \, E/C$, where E is the known energy input. \\
The derivative of the magnetisation response measured as flux in the SQUID as a function of temperature (figure \ref{fig:magnetisation_3}) $\mathrm{d} \Phi_{\mathrm{S}} / \mathrm{d}T$ can be used as translation factor in order to calculate the temperature increase $\Delta T$ corresponding to a certain flux signal in the SQUID $\Delta \Phi_{\mathrm{S}}$ : $\Delta T \; = \; \Delta \Phi_{\mathrm{S}} \cdot 1 / (\mathrm{d} \Phi_{\mathrm{S}} / \mathrm{d} T)$.

With this, the heat capacity of a single MMC pixel is given by: 

\begin{equation}
	C \; = \; \frac{E_{\mathrm{K_{\alpha}}}}{\Delta \Phi_{\mathrm{S}}} \frac{\mathrm{d} \Phi_{\mathrm{S}}}{\mathrm{d} T}.
	\label{equation:heat_capacity_expression}
\end{equation}

\begin{figure}[h!]
	\centering
    \includegraphics[width=0.8\textwidth]{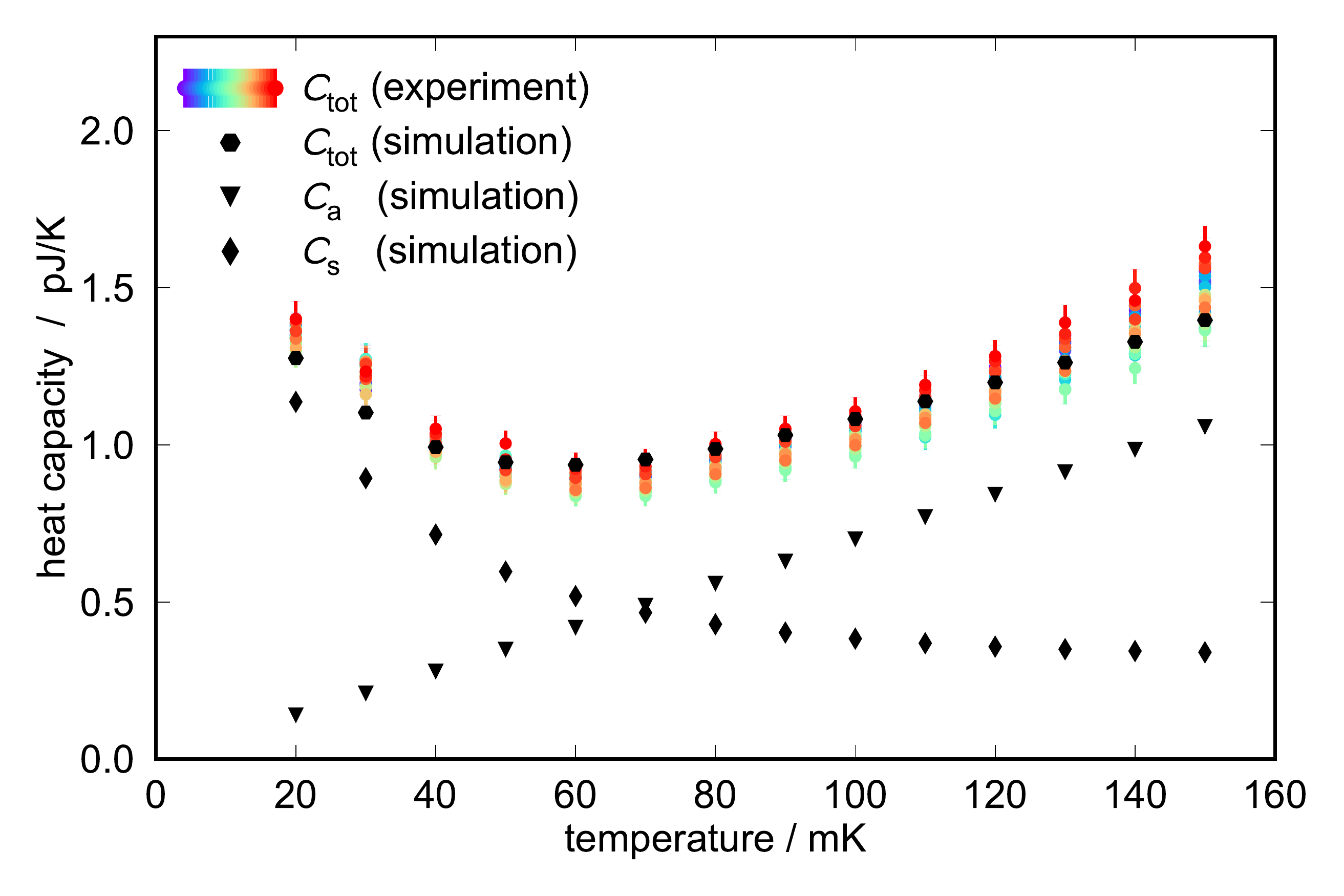} 
    \caption{The heat capacity of the single detector pixels, without implanted $^{163}\mathrm{Ho}$, deduced from experimental data with persistent current of 35 mA (coloured marks), compared with the heat capacity based on simulations (black hexagons) \cite{Herbst_simulations}. The single theoretical contributions of absorber, $C_{\mathrm{s}}$, and sensor, $C_{\mathrm{a}}$, are also plotted separately.}
	\label{fig:heatCapacity}
\end{figure}

Figure \ref{fig:heatCapacity} shows the comparison between the calculated detector heat capacity $C_{\mathrm{tot}}$, as a sum of the heat capacity contributions from the absorber $C_{\mathrm{a}}$ and from the sensor $C_{\mathrm{s}}$, and the detector heat capacity extracted from the experimental data. The absorber contribution $C_{\mathrm{a}}$ is linearly increasing with temperature, since it is dominated by the electronic term, $C_{\mathrm{a}} \, \approx \, C_{\mathrm{a,e}} \, = \, \gamma T $, where $\gamma$ is the Sommerfeld coefficient for gold. The sensor contribution $C_{\mathrm{s}}$ exhibits a $1/T^2$ dependency at temperatures above the Schottky anomaly \cite{enss_2005}.

The error bars on the experimental heat capacity values are due to the uncertainty on the flux transformer coupling, estimated to be about 4 \%, which enters in the estimation of the translation factor $\mathrm{d} \Phi_{\mathrm{S}} / \mathrm{d}T$. The theoretical expectations match with the experimental values, demonstrating the reliability of the detector model used for the optimisation process. \\
The heat capacity values show a spread for different channels, ranging from 7.6\% at 20 mK to about 14\% at 100 mK, but the overall agreement is within the expectations.

\subsubsection{Energy resolution of the ECHo-100k detector}
\label{energy_resolution}

The benchmark energy resolution for the ECHo-100k experimental phase is $5 \, \mathrm{eV}$ FWHM, and this has guided the optimisation process of the ECHo-100k detector.

\begin{figure}[h]
     	\centering
     	\begin{subfigure}[b]{0.32\textwidth}
        	 \centering
         	\includegraphics[width=\textwidth]{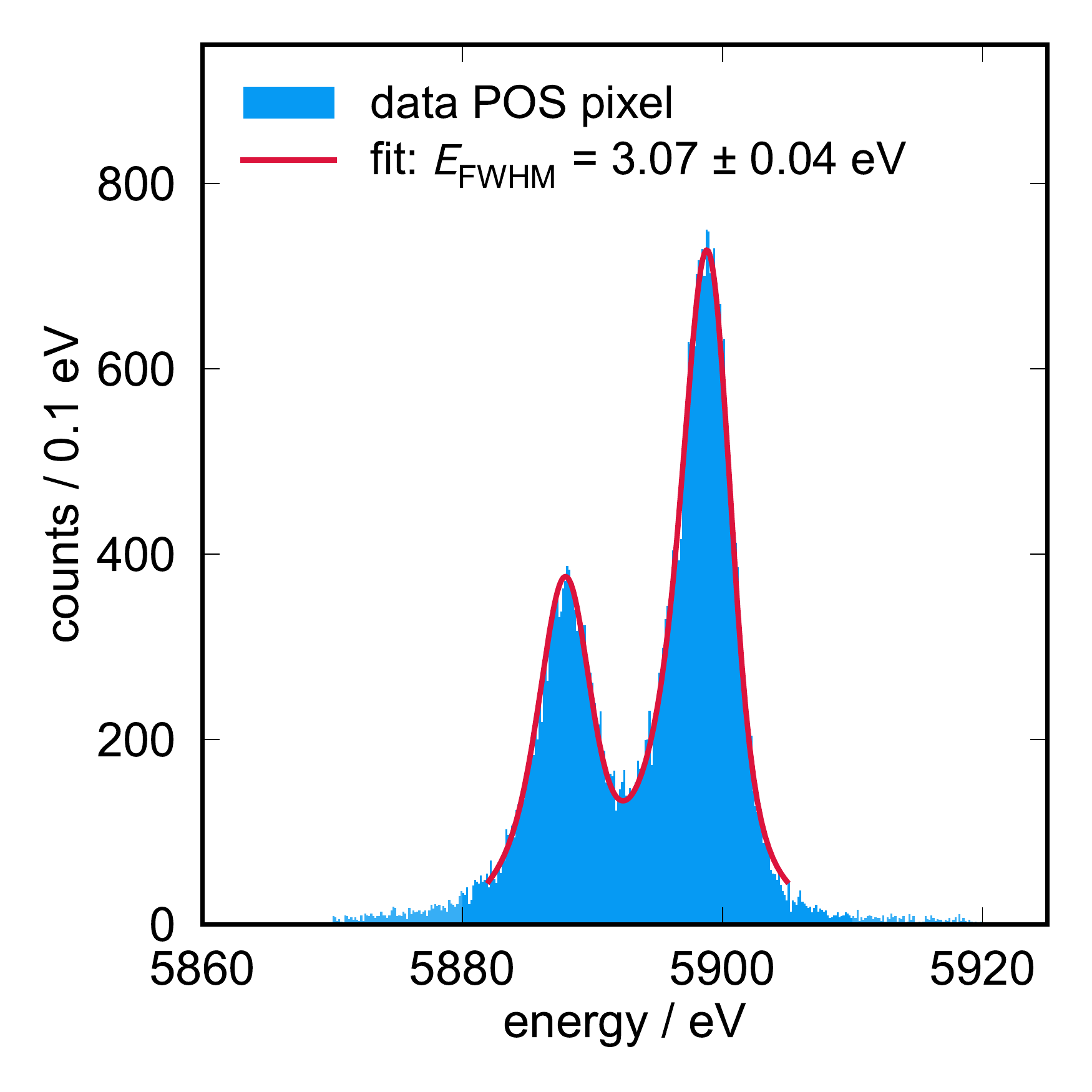}
         	\caption{}
         	\label{subfig:energyResolution_channel_17_POSP}
     	\end{subfigure}
     	\begin{subfigure}[b]{0.32\textwidth}
        	 \centering
         	\includegraphics[width=\textwidth]{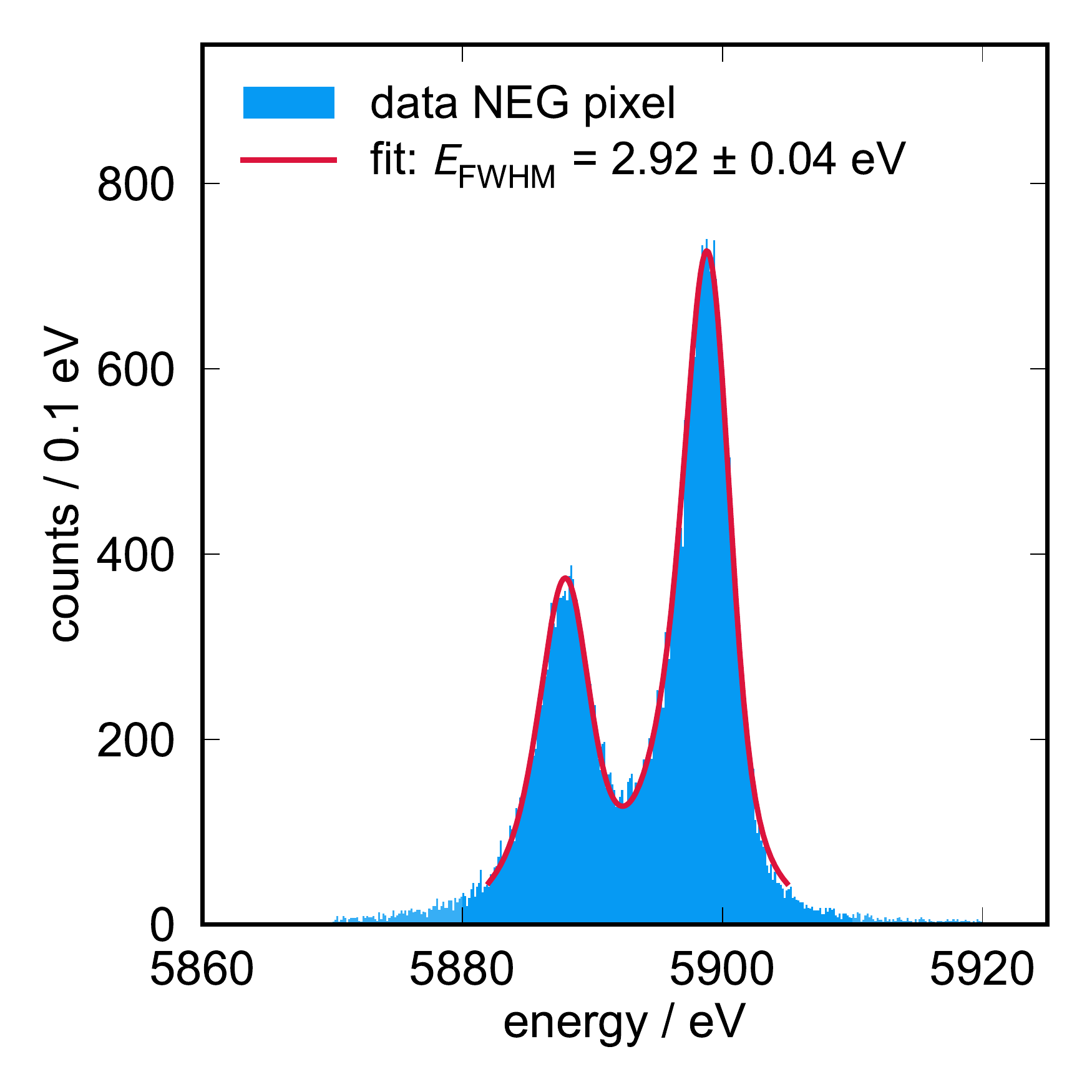}
         	\caption{}
         	\label{subfig:energyResolution_channel_17_NEGP}
     	\end{subfigure}
     	\begin{subfigure}[b]{0.32\textwidth}
        	 \centering
         	\includegraphics[width=\textwidth]{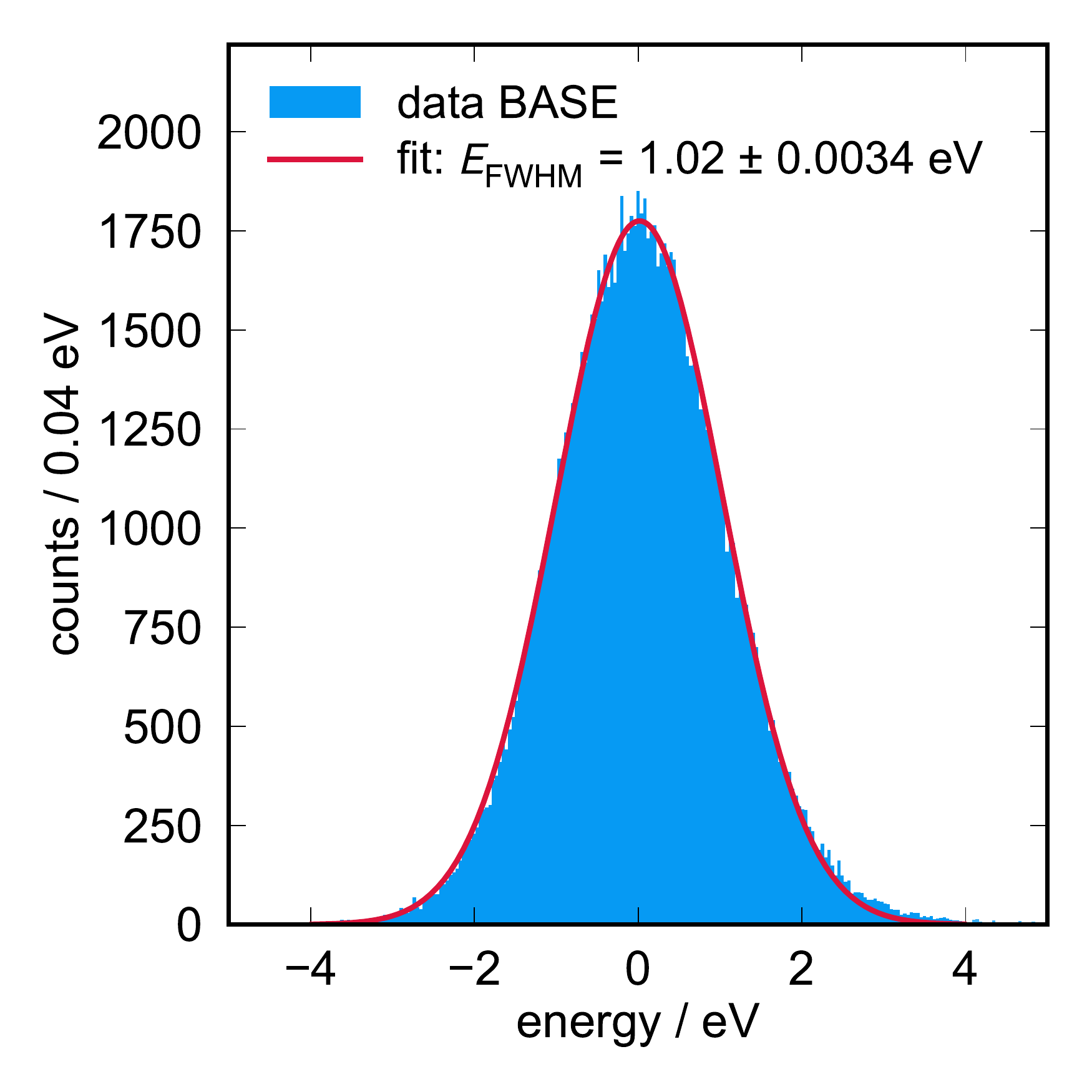}
         	\caption{}
         	\label{subfig:energyResolution_channel_17_BASE}
     	\end{subfigure}
        	
	\caption{$\mathrm{K}_{\alpha}$ doublet of the $^{55}\mathrm{Fe}$ calibration source, measured with the detector pixel not implanted with $^{163}\mathrm{Ho}$, with positive polarity signals labelled as "POS" in {\bf (a)}, and with the detector pixel with negative polarity signals labelled as "NEG" in {\bf (b)}. 
	Histogram of baseline traces with a Gaussian fit is shown in {\bf (c)}.
	}
	\label{fig:energyResolution_channel_17}
\end{figure}

\begin{figure}[h!]
	\centering
    \includegraphics[width=0.8\textwidth]{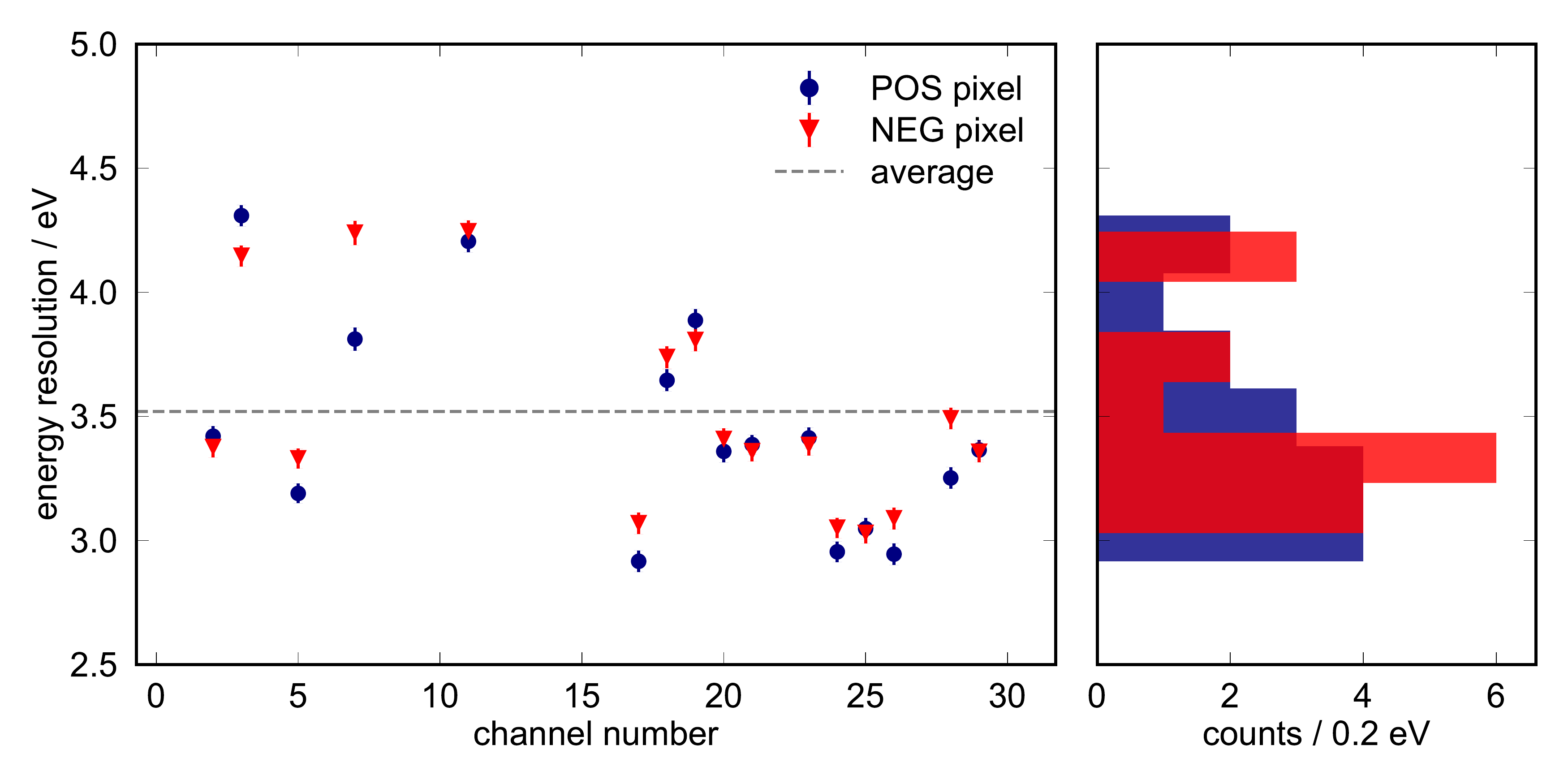} 
	\caption{The energy resolution $\Delta E_{\mathrm{FWHM}}$ from the fit of the $\mathrm{K}_{\alpha}$ doublet for each measured detector pixel of the ECHo-100k detector. The average value is indicated with a grey dashed line. The channel number on the x-axis correspond to the on-chip channel labels.}
	\label{fig:energyResolution_allChannels}
\end{figure}

An $^{55}\mathrm{Fe}$ calibration source is used to determine the achieved energy resolution in the keV energy range, by measuring the detector response corresponding to the $^{55}\mathrm{Fe}$ $\mathrm{K}_{\alpha}$ line, which exhibits a hyperfine splitting of $12 \, \mathrm{eV}$.
The measured $^{55}\mathrm{Fe}$ $\mathrm{K}_{\alpha}$ doublet is fit with a convolution of two contributions, namely the natural line-shape which is modelled as a sum of Lorentzians \cite{K_alpha_Hoelzer1997}, and the detector response which is modelled by a Gaussian function. The full width half maximum (FWHM) of the Gaussian function is left as a free parameter in the fit and it corresponds to the detector energy resolution. Figure \ref{fig:energyResolution_channel_17} shows an example fit for two detector pixels belonging to the same channel, while the energy resolutions of all measured pixels\footnote{Only the detector channels that have been measured are shown. Some channels could not be measured because of not working front-end SQUIDs or because of high noise level.} are summarised in figure \ref{fig:energyResolution_allChannels}.

The intrinsic energy resolution depends on the thermodynamic properties of the detector, such as the detector heat capacity, and on the operational temperature.
In reality, the energy resolution is also limited by the noise level of the single read-out channel and on the accuracy of the applied correction for temperature drifts. Due to the gradiometric layout of the detector channels, the read-out noise that affects two pixels belonging to the same channel is the same. However, the energy resolution of two pixels belonging to one channel is not always equal, as shown in figure \ref{fig:energyResolution_allChannels}, most likely due to the presence of spurious signals in the $^{55}\mathrm{Fe}$ spectrum (e.g.~sensor hits caused by a damage in the absorber) or because of a slightly different heat capacity of the two pixels due to non identical sensor and absorber volumes. 
In addition, the reduced thickness of the individual layers with respect to previous detector design \cite{ECHo-1k} makes the overall effect of the yet small differences in the microfabrication of the pixels more pronounced. 
The average energy resolution at $6 \, \mathrm{keV}$ of all measured pixels is $3.52 \, \mathrm{eV}$ FWHM with a standard deviation of $0.41 \, \mathrm{eV}$. The best energy resolution achieved in this measurement is $2.92 \, \pm \, 0.04$ eV FWHM, and corresponds to the detector pixel shown in figure \ref{subfig:energyResolution_channel_17_NEGP}.

Figure \ref{subfig:energyResolution_channel_17_BASE} shows the baseline energy resolution, i.e.~the energy resolution extracted from the Gaussian distribution of untriggered noise traces, for one read-out channel.
For all the read-out channels characterised by the best noise level, which is estimated to be about $0.3 \, \mathrm{\Phi_0/\sqrt{Hz}}$, the resulting baseline energy resolution is close to $1.0 \, \mathrm{eV}$. 
This value is in a fairly good in agreement with the expectations from dedicated simulations \cite{Herbst_simulations}. 
Taking into account all the pixels, the resulting averaged baseline resolution is $1.11 \, \mathrm{eV}$ with a standard deviation of $0.21 \, \mathrm{eV}$. 
The degradation of the energy resolution obtained with data from $\mathrm{K}_{\alpha}$ photon events at $6 \, \mathrm{keV}$ with respect to the resulting baseline energy resolution can be explained by a non perfected correction of the remaining temperature variations.

\section{Summary and outlook}
\label{SEC:conclusion}

The first phase of the ECHo experiment, ECHo-1k, successfully demonstrated the feasibility of the microfabrication of MMC arrays implanted with $^{163}$Ho \cite{ECHo-1k}. The results of this phase have paved the way towards the next phase of the ECHo experiment, ECHo-100k. 
For the ECHo-100k phase, the requirements in terms of number of detector pixels, $^{163}$Ho activity per pixel and energy resolution become substantially more stringent: from 60 pixels to 12000 pixels, from $1 \, \mathrm{Bq}$/pixel to $\sim 10 \, \mathrm{Bq}$/pixel and from $10 \, \mathrm{eV}$ FWHM to $5 \, \mathrm{eV}$ FWHM. 
In order to satisfy these challenging requirements, a new detector design has been developed and implemented for the ECHo-100k experimental phase.
The key improvements of the ECHo-100k design concern the energy resolution, the pulse shape, the quantum efficiency and the upgrade to the multi-chip operation via multiplexed read-out scheme  \cite{MUX_2018} \cite{MUX_2019} \cite{Richter2021} \cite{Ahrens2022}. 
The ECHo-100k detectors have been successfully fabricated, exploiting micro-lithographic techniques, and fully characterised, showing excellent performance and demonstrating an outstanding energy resolution, reaching 3\,eV FWHM at 6\,keV. 

The next steps will be to verify the detector response with the final $^{163}$Ho activity of about 10\,Bq/pixel, and proceed with the production and implantation of about 12000 detector pixels, i.e.~about 200 detector chips. 

The re\-sul\-ting high-resolution $^{163}$Ho spectrum with about $10^{13}$ electron capture events will allow to set a limit below $2 \, \mathrm{eV}$ on the effective electron neutrino mass. 

\section*{Acknowledgements}
This work has been performed in the framework of the DFG Research Unit FOR2202 “Neutrino Mass Determination by Electron Capture in 163Ho, ECHo” (funding under DU 1334/1-1 and DU 1334/1-2, EN 299/7-1 and EN 299/7-2, EN 299/8-1, GA 2219/2-1 and GA 2219/2- 2). F. Mantegazzini acknowledges support by the Research Training Group HighRR (GRK 2058) funded through the Deutsche Forschungsgemeinschaft, DFG. The $^{163}$Ho chemical separation and implantation have been performed at the Department of Chemistry of Johannes Gutenberg University Mainz and at the RISIKO facility, Johannes Gutenberg University Mainz, respectively: we acknowledge the precious work by Holger Dorrer, Christopher E.~Düllmann, Nina Kneip, Tom Kieck and Klaus Wendt. Finally, we thank the cleanroom team at the Kirchhoff-Institute for Physics for technical support during device fabrication.

\section{Appendix}
\label{SEC:appendix}

\subsection*{Microfabrication steps}

\begin{figure}[h!]
    \centering
    \begin{subfigure}[t]{0.3\textwidth}
        \centering
        \includegraphics[width=\linewidth]{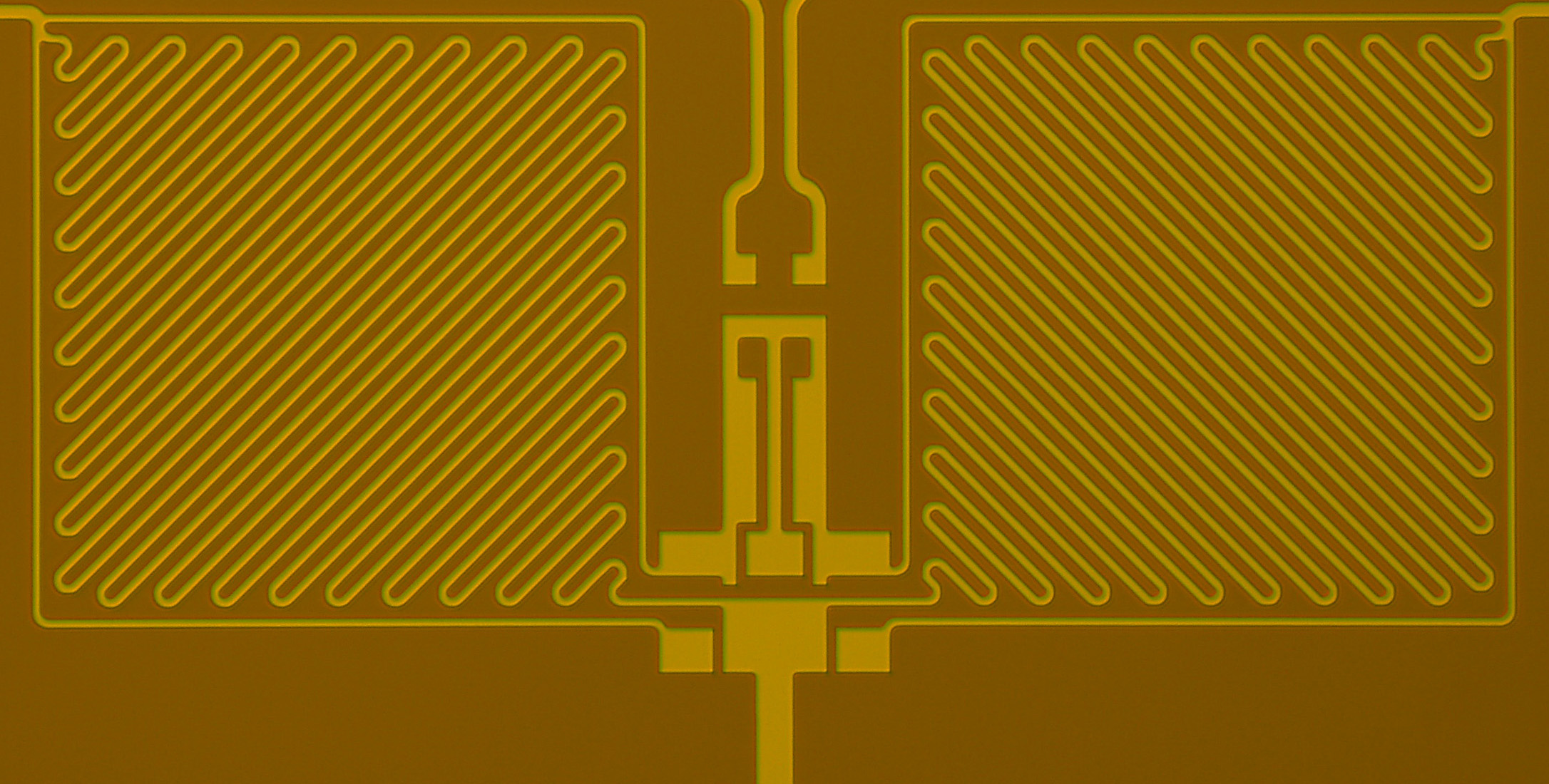}
        \caption{} \label{SUBFIG:ECHo-100k_l1}
    \end{subfigure}
    \hfill
    \begin{subfigure}[t]{0.3\textwidth}
        \centering
        \includegraphics[width=\linewidth]{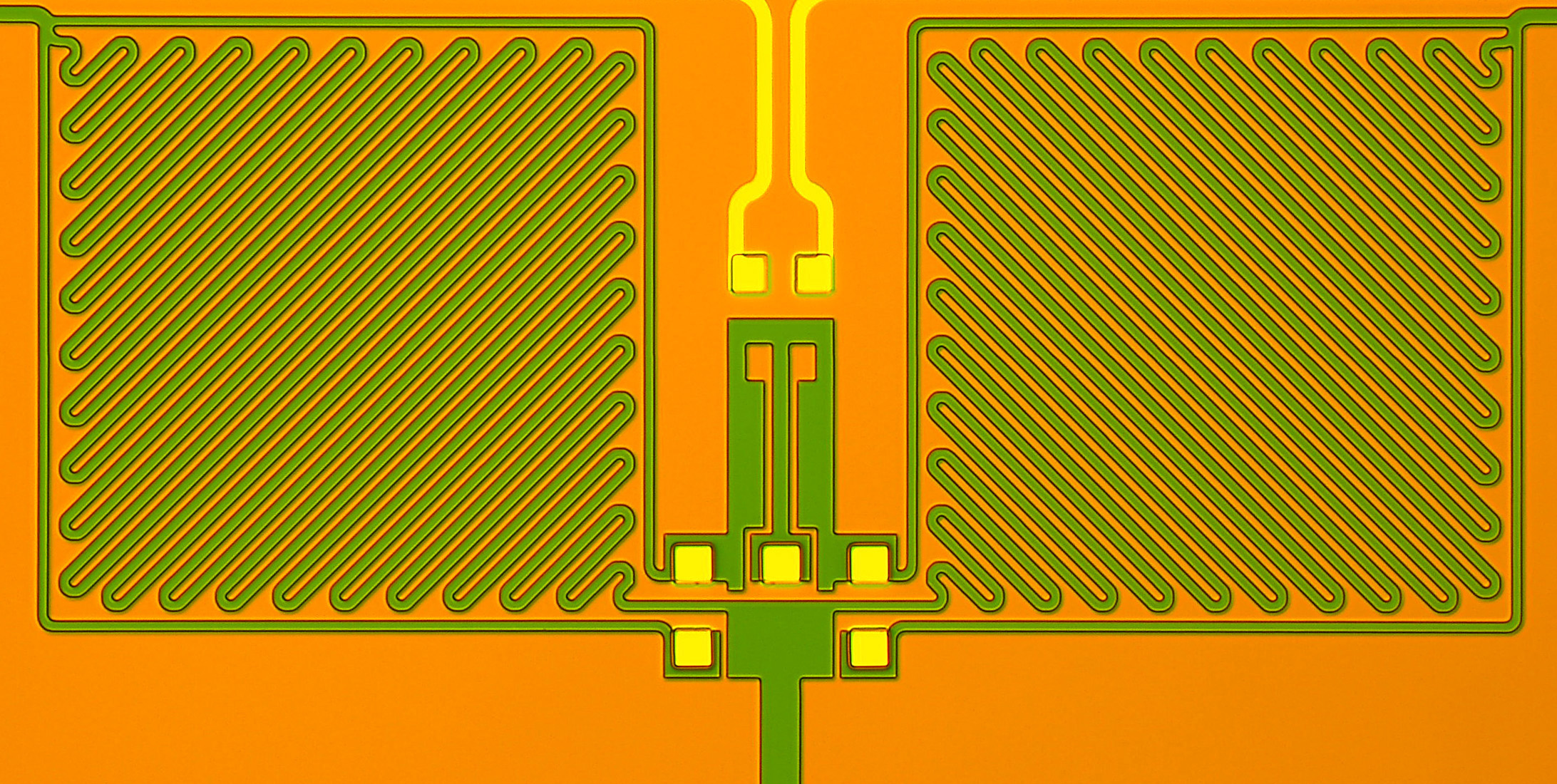}
        \caption{} \label{SUBFIG:ECHo-100k_l2}
    \end{subfigure}
    \hfill
    \begin{subfigure}[t]{0.3\textwidth}
        \centering
        \includegraphics[width=\linewidth]{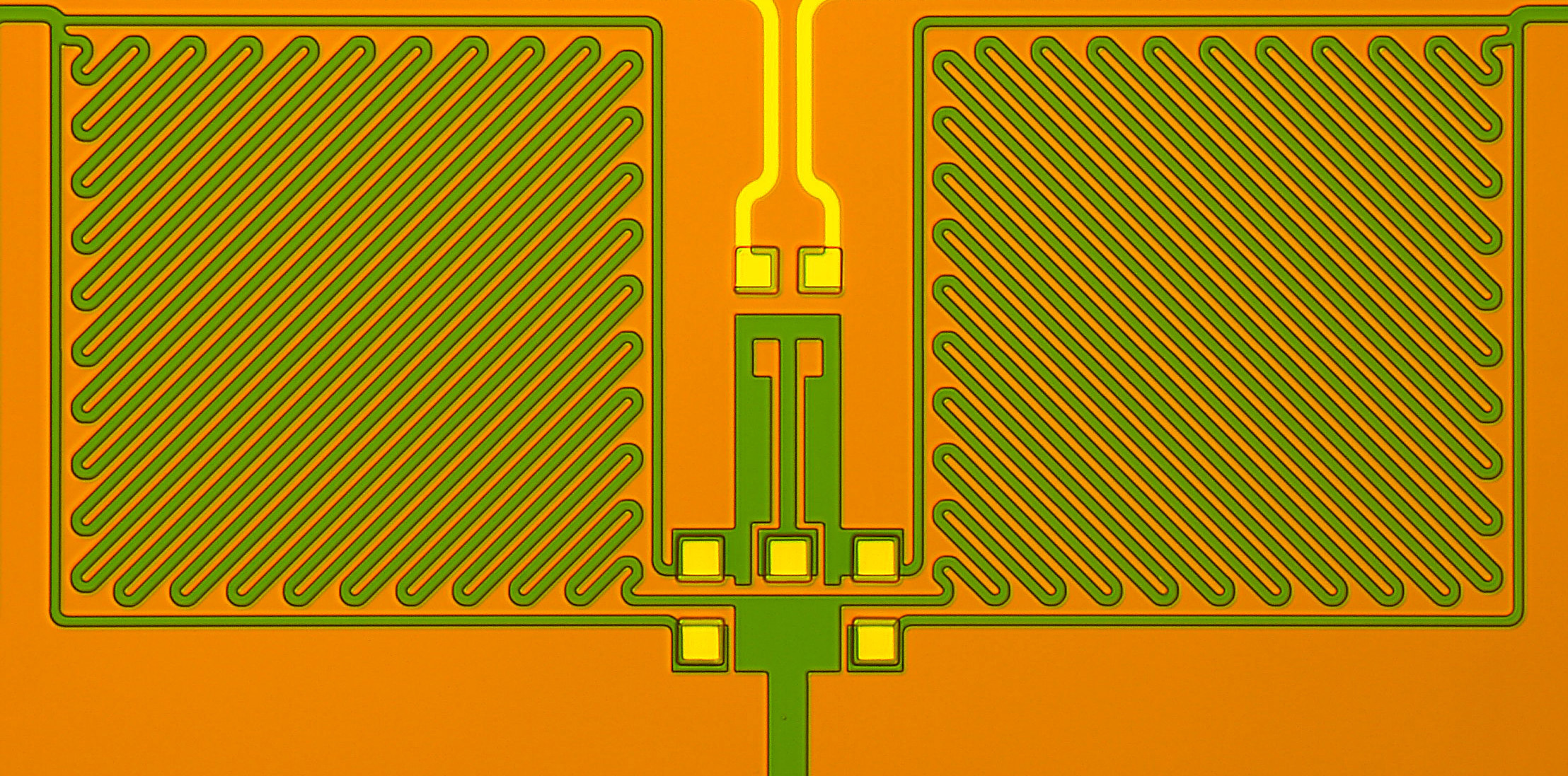}
        \caption{} \label{SUBFIG:ECHo-100k_l3}
    \end{subfigure}
    \hfill
    \begin{subfigure}[t]{0.3\textwidth}
        \centering
        \includegraphics[width=\linewidth]{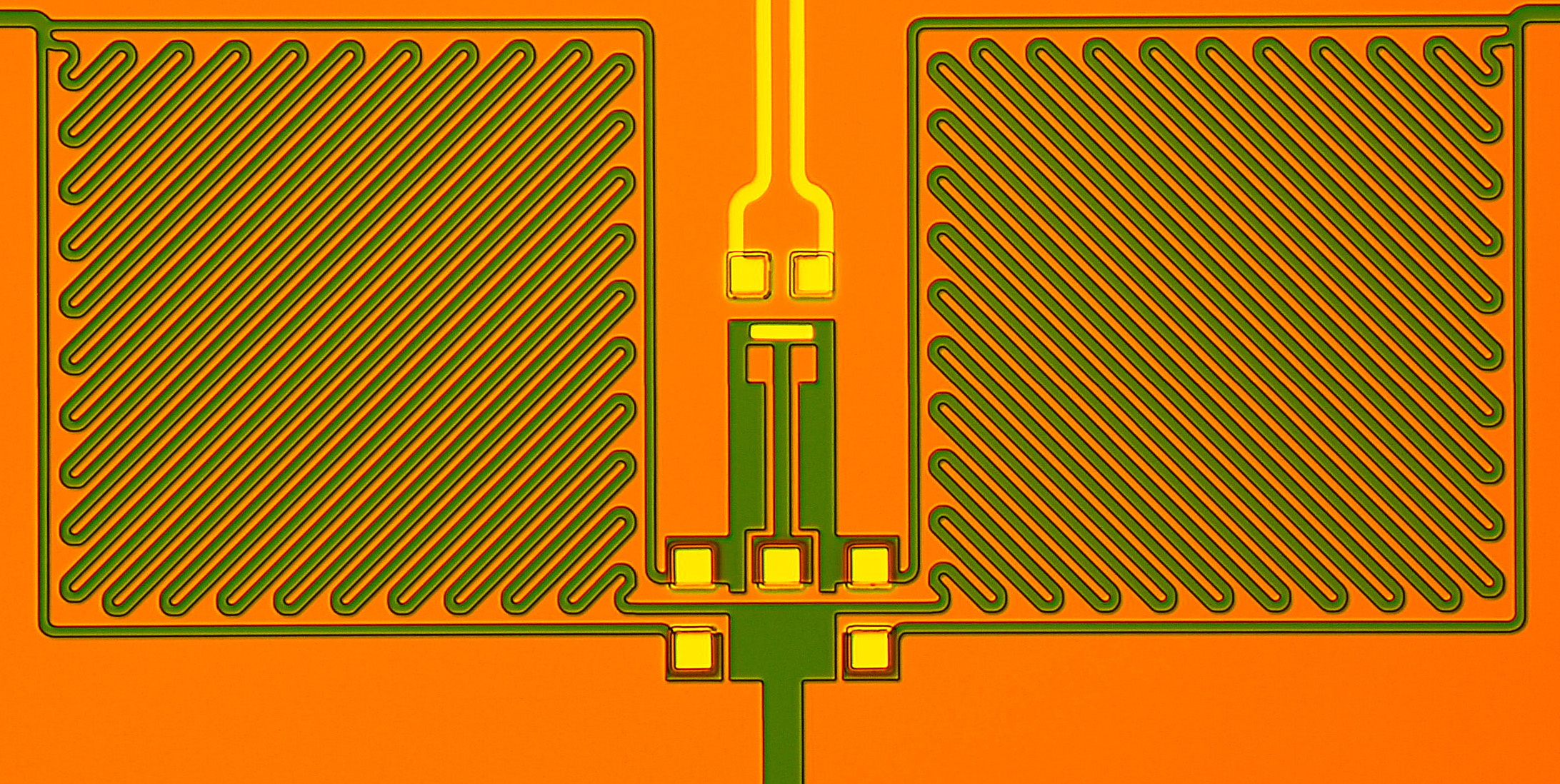}
        \caption{} \label{SUBFIG:ECHo-100k_l4}
    \end{subfigure}
    \hfill
        \begin{subfigure}[t]{0.3\textwidth}
        \centering
        \includegraphics[width=\linewidth]{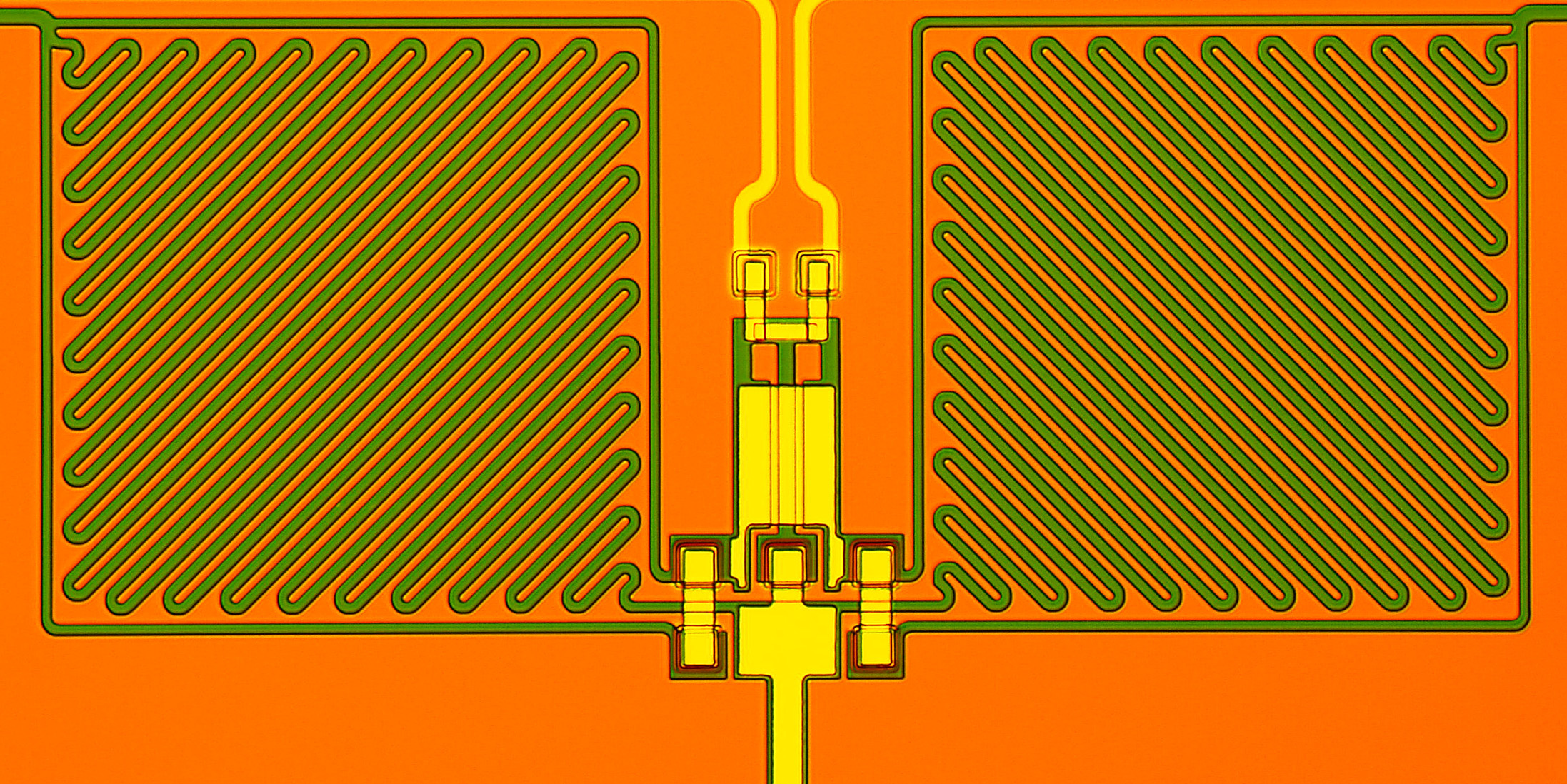}
        \caption{} \label{SUBFIG:ECHo-100k_l5}
    \end{subfigure}
    \hfill
    \begin{subfigure}[t]{0.3\textwidth}
        \centering
        \includegraphics[width=\linewidth]{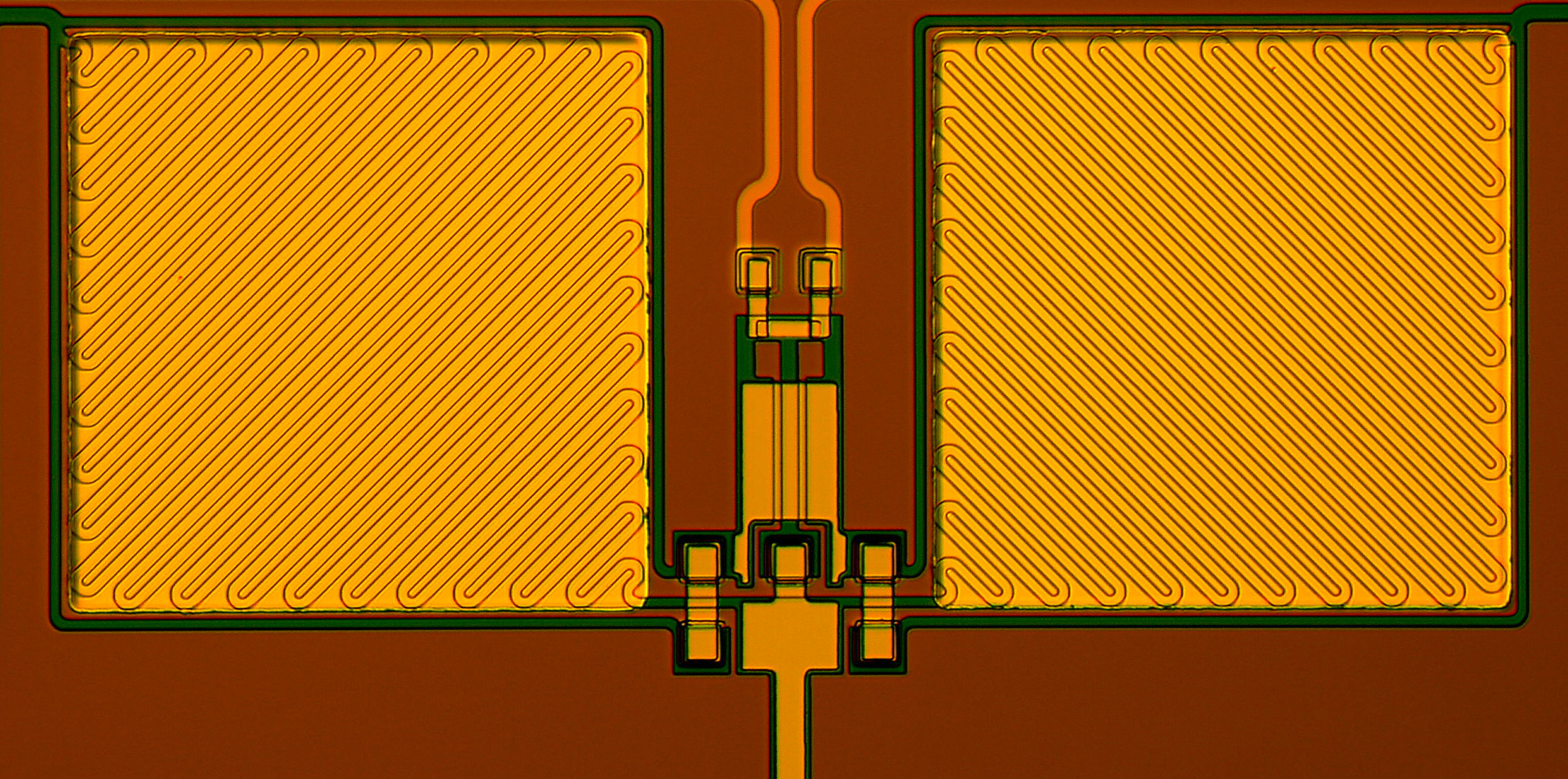}
        \caption{} \label{SUBFIG:ECHo-100k_l6}
    \end{subfigure}
    \hfill
        \begin{subfigure}[t]{0.3\textwidth}
        \centering
        \includegraphics[width=\linewidth]{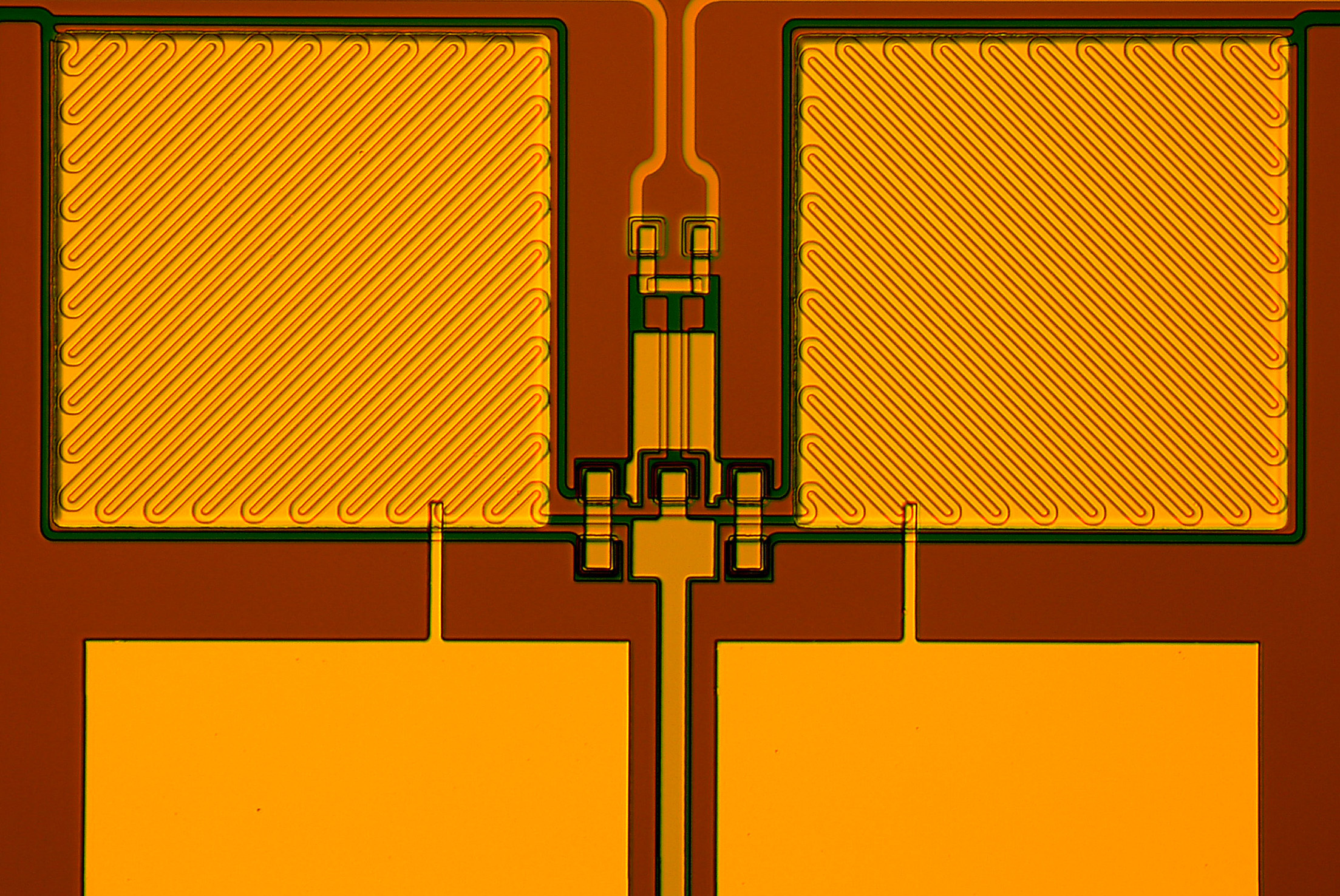}
        \caption{} \label{SUBFIG:ECHo-100k_l7}
    \end{subfigure}
    \hfill
    \begin{subfigure}[t]{0.3\textwidth}
        \centering
        \includegraphics[width=\linewidth]{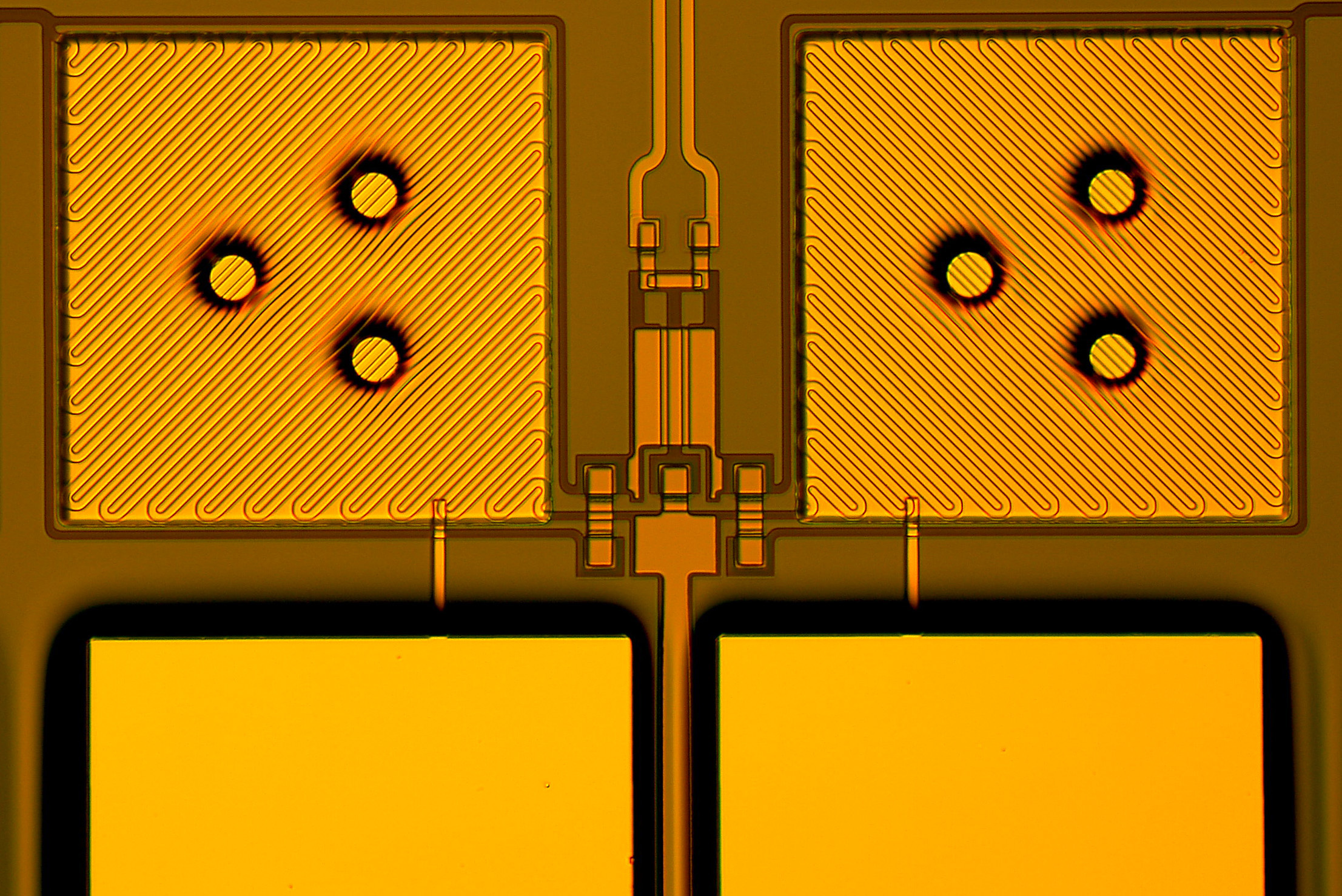}
        \caption{} \label{SUBFIG:ECHo-100k_l8}
    \end{subfigure}
    \hfill
    \begin{subfigure}[t]{0.3\textwidth}
        \centering
        \includegraphics[width=\linewidth]{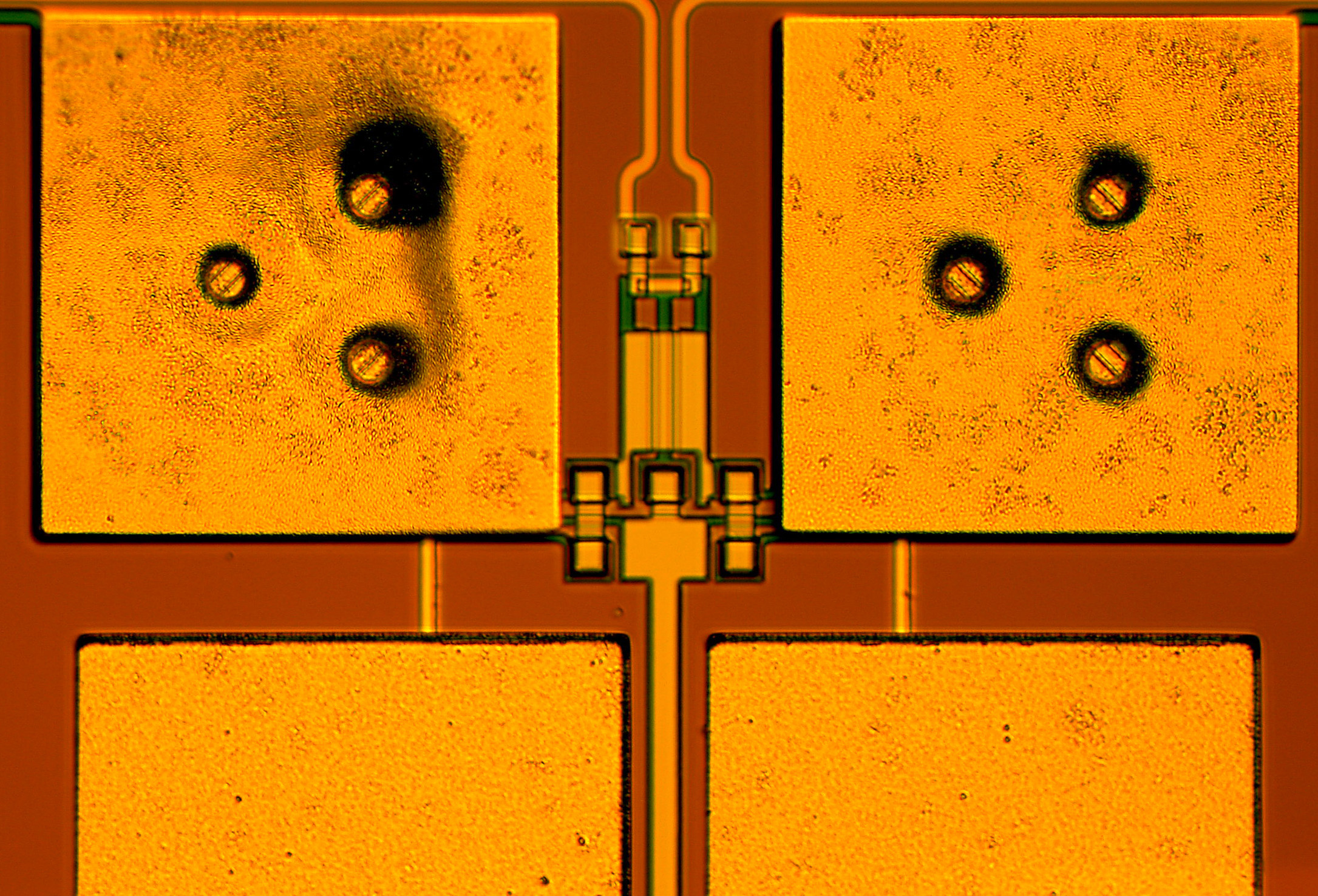}
        \caption{} \label{SUBFIG:ECHo-100k_l9}
    \end{subfigure}

    \caption{Microscope pictures of the fabrication steps of the ECHo-100k detector. \textbf{(a)} first niobium layer, \textbf{(b)} anodisation + first silicon dioxide isolation, \textbf{(c)} second silicon dioxide isolation, \textbf{(d)} gold-palladium heater switch, \textbf{(e)} second niobium layer, \textbf{(f)} silver-erbium sensor, \textbf{(g)} gold thermalisation layer, \textbf{(h)} gold pillars, \textbf{(i)} gold first absorber layer. The colours change from layer 1 to layer 2 due to the isolation deposition and the anodisation process. The colours of the remaining layers are affected by the illumination and the presence of photoresist.}
    \label{FIG:ECHo-100k_fabrication_steps}
\end{figure}

The fabrication of the ECHo-100k detectors has been performed in the cleanroom facility of the Kirchhoff-Institute for Physics, Heidelberg University. Table \ref{TAB:fabrication} reports the microfabrication steps, while in figure \ref{FIG:ECHo-100k_fabrication_steps} the microscope pictures of each steps are shown.

\begin{table}[h!]
\centering
\begin{tabular}{ |p{0.4cm}|p{4cm}|p{1.4cm}|p{1.6cm}|p{3.9cm}|  }
 \hline
 \multicolumn{5}{|c|}{Fabrication steps} \\
 \hline
 \hline 
 \# & Layer & Material & Thickness & Deposition technique\\
 \hline
 1 & \makecell[l]{Meanders, \\ SQUID lines (layer 1)}  & Nb & 250 nm & Sputtering + etching\\
 2 & Isolation  & $\mathrm{Nb_{2}O_{5}}$ & - & Anodisation\\
 3 & Isolation  & $\mathrm{SiO_{2}}$ & 175 nm & Sputtering + lift-off\\
 4 & Isolation  & $\mathrm{SiO_{2}}$ & 175 nm & Sputtering + lift-off\\
 5 & Heaters  & AuPd & 150 nm & Sputtering\\
 6 & SQUID lines (layer 2)  & Nb & 600 nm & Sputtering + lift-off\\
 7 & Sensor  & AgEr & 480 nm & Sputtering + lift-off\\
 8 & Thermalisation  & Au & 300 nm & Sputtering + lift-off\\
 9 & Pillars  & Au & 100 nm & Sputtering\\
 10 & Absorber - 1st layer & Au & 3 $\mathrm{\upmu m}$ & Electroplating + lift-off\\
 11 & $^{163}$Ho host material & Ag & 100 nm & Sputtering\\
 12 & $^{163}$Ho implantation & $^{163}$Ho & - & Ion-implantation\\
 13 & $^{163}$Ho host material & Ag & 100 nm & Sputtering\\
 14 & Absorber - 2nd layer & Au & 3 $\mathrm{\upmu m}$ & Sputtering + lift-off\\
 \hline
\end{tabular}
\caption{Microfabrication steps of the ECHo-100k detector chip.}
\label{TAB:fabrication}
\end{table}

\newpage
\subsection*{Measurement of the inductance of the pick-up coil}

\begin{figure}[h!] 
	\centering
	\includegraphics[width=.55\textwidth]{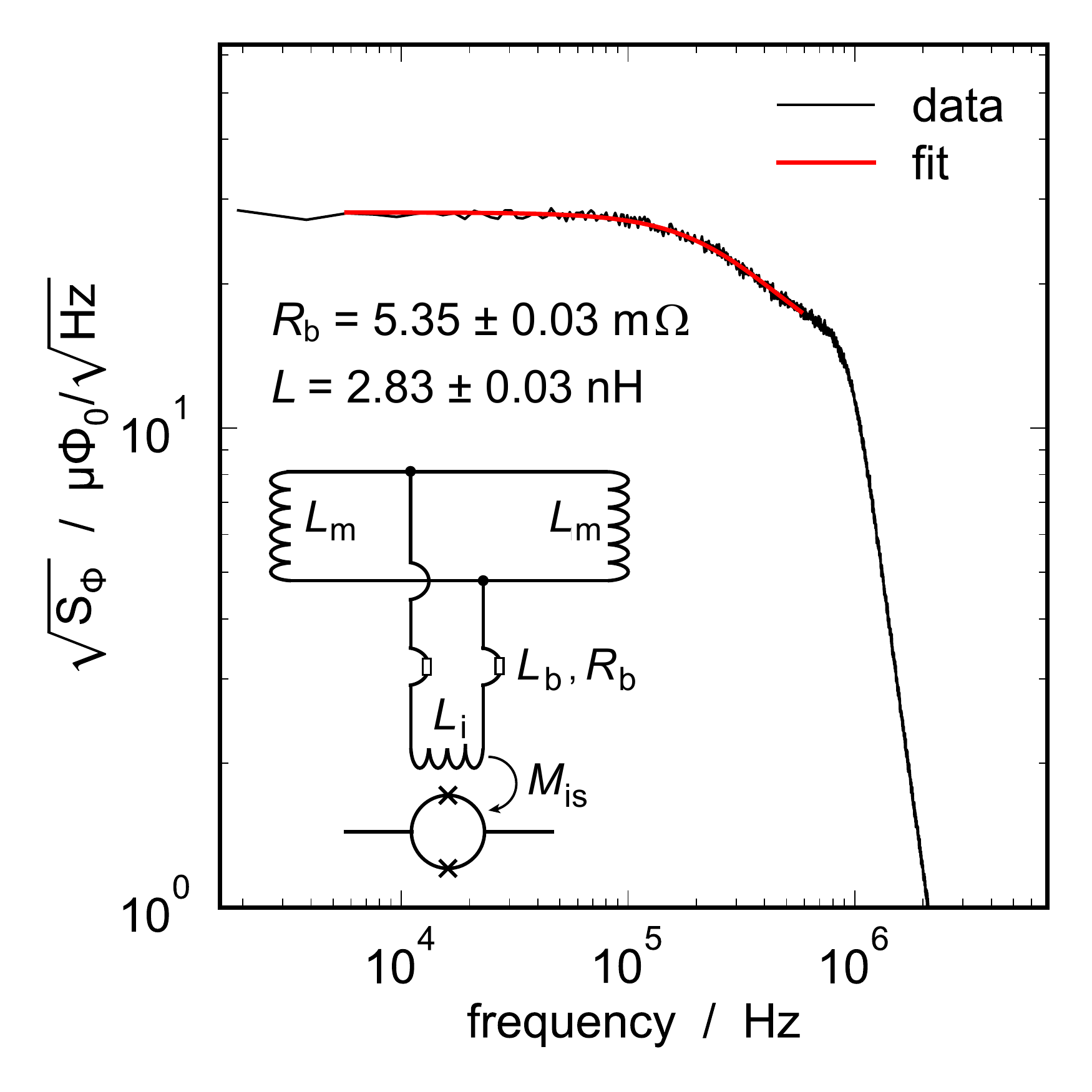}
	\caption{Measured spectral flux density in the SQUID for an ECHo-100k detector channel. The data is shown in black while the fit with the parameters $R_\mathrm{b}$ and $L$ quoted in the graphic is shown in red. The inset shows the layout of the corresponding circuit. } 
	\label{FIG:noise_ECHo-100k}
\end{figure}

The inductance $L$ of the circuit formed by the input coil of the SQUID and the detector pick-up coil together with the resistance of the normal-conducting bonding wires $R_\mathrm{b}$ constitute a low-pass filter with a cut-off frequency $f_\mathrm{c} = R_\mathrm{b}/(2 \pi L) $. The corresponding circuit is shown in the inset of figure \ref{FIG:noise_ECHo-100k}. 
The expected total noise spectral density is given by

\begin{equation}
    S_{\mathrm{\Phi_\mathrm{S}}} = M_{\mathrm{is}}^2 \cdot 4 k_\mathrm{B} T / R_\mathrm{b} \cdot 1 / (1+(f/f_\mathrm{c})^2) + S_{\mathrm{\Phi_\mathrm{S}}}^\mathrm{S,w}
\end{equation}

\noindent where $M_\mathrm{is}$ is the mutual inductance between the pick-up coil and the SQUID and $S_\mathrm{\Phi_\mathrm{S}}^\mathrm{S,w}$ is the white noise of the SQUID.

The spectral flux density measured in the SQUID, with a ECHo-100k detector channel connected to it, is shown in figure \ref{FIG:noise_ECHo-100k}, with the corresponding numerical fit. \\

The inductance of a single meander-shaped pick-up coil is given by $L_\mathrm{m} = 2 \cdot (L - L_\mathrm{i} - L_\mathrm{b})$, where $L = 2.83 \pm 0.03 \, \mathrm{nH}$ is the resulting total inductance given by the fit, $L_\mathrm{i} = 1.24 \, \mathrm{nH}$ is the inductance of the SQUID input coil, $L_\mathrm{b} \approx c \cdot R_\mathrm{b} $ is the inductance of the bonding wires, with the parameter $c = 0.10 \pm 0.02 \, \mathrm{nH/m \Omega}$ experimentally determined and $R_\mathrm{b} = 5.35 \pm 0.03 \, \mathrm{m\Omega}$ from the fit.
The derived value for the pick-up coil inductance $L_\mathrm{m} = 2.1 \pm 0.2 \, \mathrm{nH}$ is consistent with the expected value.

\subsection*{Methodology for the magnetisation measurement}

Two ECHo-100k detector channels are non-gradiometric by design, and are dedicated to temperature monitoring. These channels feature one normal, non-implanted, MMC pixel, and a naked meander-shaped pick-up coil in the place where the other pixel should normally be. This makes the baseline offset of these channels sensitive to small fluctuations in the detector temperature, and as such can be employed for temperature monitoring and correction.

Due to this non-gradiometric design, temperature channels can also be employed for measuring temperature dependence of the magnetisation response of the sensor. As only one of the two MMC pixels features a sensor, a change of the baseline value following a change in temperature is directly proportional to the change in the magnetisation. 

The magnetisation measurement consists of multiple predefined temperature steps, usually 10 to 30 of them, in between which the detector is allowed to properly thermalise. The thermalisation time, usually 30 to 90 minutes, was introduced in order to ensure that the temperature of the detector is the same as the recorded temperature of the mixing chamber plate.
As the acquisition of the SQUID voltage output and the temperature of the mixing chamber are synchronised and measured continuously, the resulting curves of magnetisation as a function of time and temperature as a function of time will feature plateaus, each corresponding to one temperature step, and with the length corresponding to selected thermalisation time. \\

\begin{figure}[h!]
     	\centering
     	\begin{subfigure}[b]{0.32\textwidth}
        	 	\centering
         	\includegraphics[width=\textwidth]{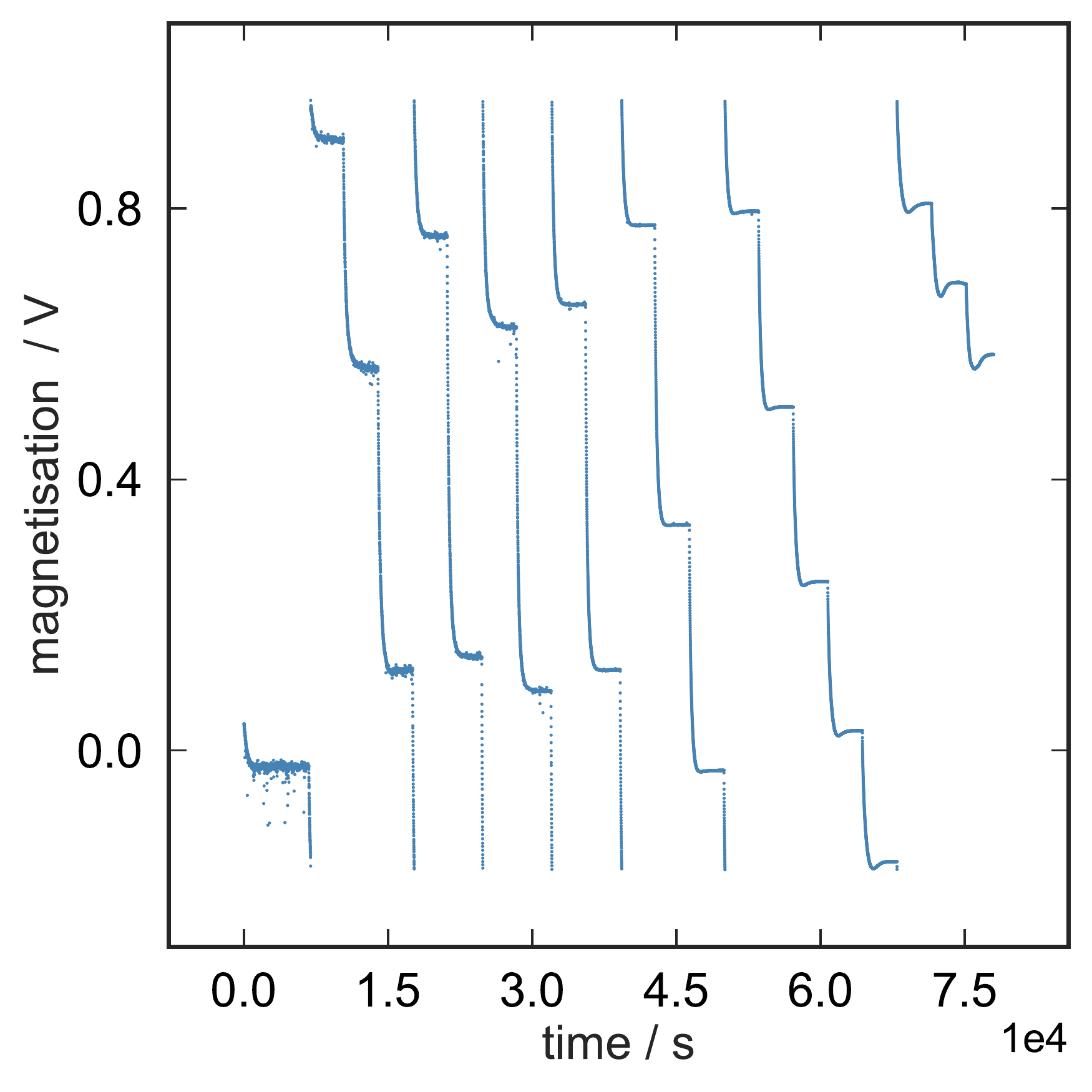}
         	\caption{}
         	\label{subfig:magnetisationJumpsChange}
     	\end{subfigure}
     	\begin{subfigure}[b]{0.32\textwidth}
        	 	\centering
         	\includegraphics[width=\textwidth]{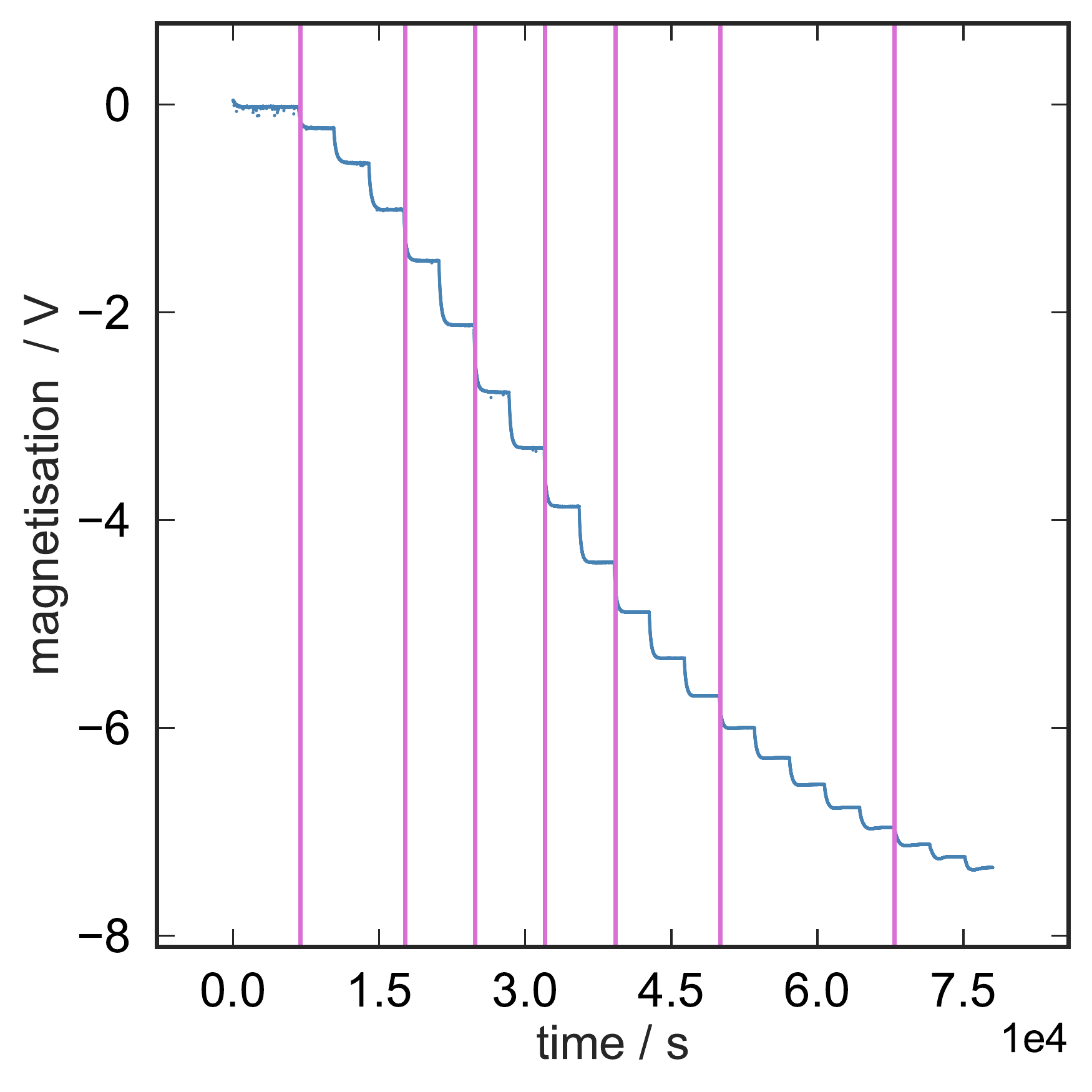}
         	\caption{}
         	\label{subfig:magnetisationJumpsCorrected}
     	\end{subfigure}
     	\begin{subfigure}[b]{0.32\textwidth}
         	\centering
         	\includegraphics[width=\textwidth]{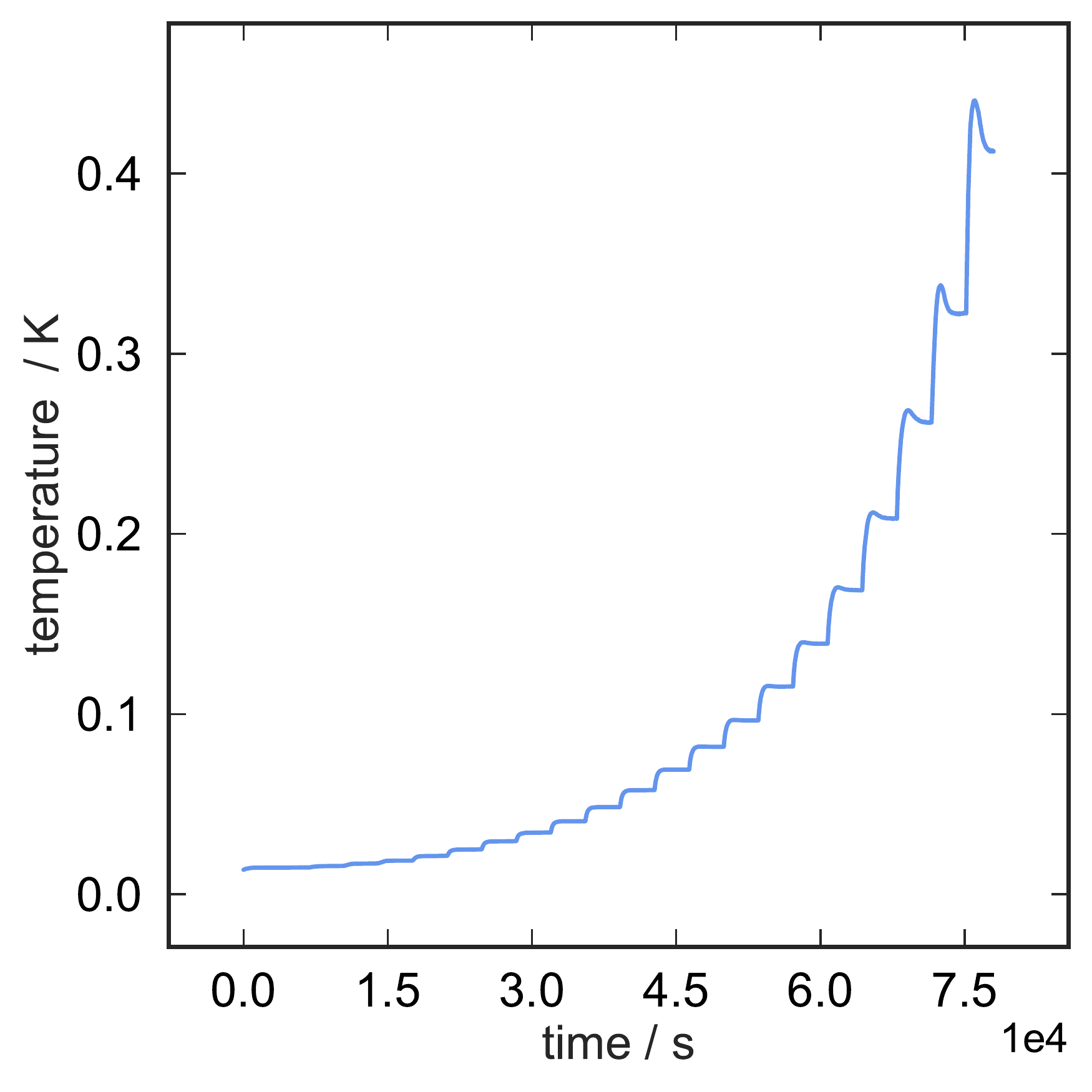}
         	\caption{}
         	\label{subfig:temperatureChange}
     	\end{subfigure}
        	
	\caption{Recorded data during the magnetisation measurement: {\bf (a)} voltage signal output of the SQUID, showing jumps due to SQUID reset, {\bf (b)} voltage signal output of the SQUID, after correcting for individual jumps, {\bf (c)} recorded temperature of the mixing chamber plate of the dilution refrigerator. The chosen value of $R_{\mathrm{fb}}$ used during the measurement was $\mathrm{10 \, k\Omega}$, and the extracted $V/\Phi_0$ is $0.377 \, \mathrm{V/\Phi_0}$.}
    \label{fig:magnetisation_measurement}
\end{figure}

In the flux locked loop (FLL) configuration \cite{SQUID_handbook}, the voltage signal $V$ and the corresponding flux in the SQUID $\Phi_{\mathrm{S}}$ are related by $\Phi_{\mathrm{S}} \, = \, - V \cdot M_{\mathrm{fb}}/R_{\mathrm{fb}}$, where $M_{\mathrm{fb}}$ is the mutual inductance between the front-end SQUID and the feedback coil, $R_{\mathrm{fb}}$ is the feedback resistance and $V$ is the voltage signal across the feedback resistance. The value of $M_{\mathrm{fb}}$ is determined by the design of the used SQUID, while $R_{\mathrm{fb}}$ is set to a desired value via the SQUID software control\footnote{Typical $R_{\mathrm{fb}}$ values are between $\mathrm{10 \, k\Omega}$ and $\mathrm{100 \, k\Omega}$ and they are set with the SQUIDViewer Sofwtare, Magnicon \url{http://www.magnicon.com/squid-electronics/squidviewer-software}.} prior to the beginning of the measurement. 
The choice of $R_{\mathrm{fb}}$ together with the value of $M_{\mathrm{fb}}$ defines the voltage corresponding to a change of flux, equal to $\Phi_0$ in the SQUID. 
Considering that the voltage range of the ADC is limited, an automatic reset of the SQUID is applied\footnote{When the SQUID output voltage exceeds the threshold value set by the user, in the measurements discussed here -0.2\,V, the SQUID is set in open loop and then locked back to FLL mode.} in order to keep the voltage output of the SQUID within the ADC range. This reset will result in jumps in the recorded voltage data, as can be seen in figure \ref{subfig:magnetisationJumpsChange}, that need to be corrected for. Each jump is a multiple of the flux quantum $\Phi_0$, thus allowing the voltage to flux conversion factor $V/\Phi_0$ to be experimentally determined from the values of the jump. 
The direct determination via voltage jumps is more accurate than the calculation from the design values, as the mutual inductance can slightly deviate from one SQUID to another due to fabrication inaccuracies\footnote{The measured mutual inductance of individual SQUIDs used in the reported measurements showed a spread in values up to 2.5\%.}.
Figure \ref{subfig:magnetisationJumpsCorrected} shows the voltage curve after correcting for each jump, with the positions of the jumps indicated in pink, and the corresponding temperature data is shown in figure \ref{subfig:temperatureChange}. \\

The final result of the magnetisation measurement is a curve of temperature dependence of the magnetisation, usually plotted as magnetisation in units of $\Phi_0$ as a function of the inverse temperature (given in units of 1/K), as shown in figure \ref{fig:magnetisation_3}. This curve is obtained by averaging over last 100 to 150 points (corresponding to about 2 to 3 minutes of continuous acquisition) at the end of each plateau, both in the magnetisation and in the temperature data. 
The reason why only the ends of the plateaus enter in the final analysis is that the thermalisation of the detector to the set temperature of the mixing chamber plate is not instantaneous, but happens on a finite time scale which depends on the detector set-up and the quality of the thermal connection to the mixing chamber.      
By selecting an adequate thermalisation step, we ensure that for the points selected for the final analysis the detector is at the same temperature as the recorded temperature of the mixing chamber plate.

\bibliographystyle{unsrt}
\bibliography{biblio}

\begin{thebibliography}{10}

\bibitem{DeRujula_Lusignoli}
A.~De~Rújula and M.~Lusignoli.
\newblock {Calorimetric measurements of Holmium-163 decay as tools to determine
  the electron neutrino mass}.
\newblock {\em Phys. Lett. B}, 118:429–434, 1982.

\bibitem{ECHoGeneral}
L.~Gastaldo and the ECHo~Collaboration.
\newblock {The electron capture in $^{163}$Ho experiment - ECHo}.
\newblock {\em Eur. Phys. J. Special Topics}, 226:1623–1694, 2017.

\bibitem{KATRIN2022}
The~KATRIN Collaboration.
\newblock {Direct neutrino-mass measurement with sub-electronvolt sensitivity}.
\newblock {\em Nat. Phys.}, 18:160–166, 2022.

\bibitem{ECHo_spectrum_2019}
C.~Velte et~al.
\newblock {High-resolution and low-background $^{163}$Ho spectrum:
  interpretation of the resonance tails}.
\newblock {\em Eur. Phys. J. C}, 79:1623–1694, 2019.

\bibitem{Holmes2023}
M.~Borghesi, B.~Alpert, M.~Balata, D.~Becker, D.~Bennet, E.~Celasco,
  N.~Cerboni, M.~{De Gerone}, R.~Dressler, M.~Faverzani, M.~Fedkevych,
  E.~Ferri, J.~Fowler, G.~Gallucci, J.~Gard, F.~Gatti, A.~Giachero, G.~Hilton,
  U.~Koster, D.~Labranca, M.~Lusignoli, J.~Mates, E.~Maugeri, S.~Nisi,
  A.~Nucciotti, L.~Origo, G.~Pessina, S.~Ragazzi, C.~Reintsema, D.~Schmidt,
  D.~Schumann, D.~Swetz, J.~Ullom, and L.~Vale.
\newblock An updated overview of the holmes status.
\newblock {\em Nuclear Instruments and Methods in Physics Research Section A:
  Accelerators, Spectrometers, Detectors and Associated Equipment},
  1051:168205, 2023.

\bibitem{Fle2005}
A.~Fleischmann, C.~Enss, and G.M. Seidel.
\newblock {Metallic magnetic calorimeters}.
\newblock In C.~Enss, editor, {\em {Cryogenic Particle Detection, Topics in
  Applied Physics}}, volume~99, pages 151--216. Springer-Verlag, Berlin,
  Heidelberg, New York, 2005.

\bibitem{ECHo-1k}
F.~Mantegazzini et~al.
\newblock {Metallic magnetic calorimeter arrays for the first phase of the ECHo
  experiment}.
\newblock {\em Nucl. Instrum. Meth. A}, 1030:166406, 2022.

\bibitem{Brass_HoSpectrum_2020}
M.~Brass and M.W. Haverkort.
\newblock {Ab initio calculation of the electron capture spectrum of
  $^{163}$Ho: Auger-Meitner decay into continuum states}.
\newblock {\em New J. Phys.}, 22, 2020.

\bibitem{Kem2018}
S.~Kempf, A.~Fleischmann, L.~Gastaldo, and C.~Enss.
\newblock Physics and applications of metallic magnetic calorimeters.
\newblock {\em J. Low Temp. Phys.}, 193, 2018.

\bibitem{Gastaldo_NIMA}
L.~Gastaldo, P.~Ranitzsch, F.~von Seggern, J.-P. Porst, S.~Schäfer, C.~Pies,
  S.~Kempf, T.~Wolf, A.~Fleischmann, C.~Enss, A.~Herlert, and K.~Johnston.
\newblock {Characterization of low temperature metallic magnetic calorimeters
  having gold absorbers with implanted $^{163}$Ho ions}.
\newblock {\em Nucl. Inst. Meth. A}, 711, 2013.

\bibitem{Hassel2016}
C.~Hassel, K.~Blaum, T.~Day~Goodacre, et~al.
\newblock {Recent Results for the ECHo Experiment}.
\newblock {\em J Low Temp Phys.}, 184:910--921, 2016.

\bibitem{LTD_proceedings_2021}
M.~Griedel, F.~Mantegazzini, A.~Barth, E.~Bruer, W.~Holzmann, R.~Hammann,
  D.~Hengstler, N.~Kovac, C.~Velte, T.~Wickenh{\"a}user, A.~Fleischmann,
  C.~Enss, L.~Gastaldo, H.~Dorrer, T.~Kieck, N.~Kneip, Ch.E. D{\"u}llmann, and
  K.~Wendt.
\newblock {From ECHo-1k to ECHo-100k: Optimization of high-resolution metallic
  magnetic calorimeters with embedded $^{163}\mathrm{Ho}$ for neutrino mass
  determination}.
\newblock {\em J Low Temp Phys}, 2022.

\bibitem{MUX_2018}
M.~Wegner, N.~Karcher, O.~Kr\"omer, D.~Richter, F.~Ahrens, O.~Sander, S.~Kempf,
  M.~Weber, and C.~Enss.
\newblock {Microwave SQUID Multiplexing of Metallic Magnetic Calorimeters:
  Status of Multiplexer Performance and Room-Temperature Readout Electronics
  Development}.
\newblock {\em J. Low Temp. Phys.}, 193, 2018.

\bibitem{MUX_2019}
O.~Sander, N.~Karcher, O.~Kr\"omer, M.~Kempf, S. an~Wegner, C.~Enss, and
  M.~Weber.
\newblock {Software-Defined Radio Readout System for the ECHo Experiment}.
\newblock {\em IEEE TRANSACTIONS ON NUCLEAR SCIENCE}, 66, 2019.

\bibitem{Richter2021}
D.F. Richter.
\newblock {\em {Multikanal-Auslesung von metallischen magnetischen Kalorimetern
  mittels eines vollständigen Mikrowellen-SQUID-Multiplexer-Systems}}.
\newblock {PhD thesis}, Kirchhoff-Institut f{\"u}r Physik, Universit{\"a}t
  Heidelberg, 7 2021.

\bibitem{Ahrens2022}
F.K. Ahrens.
\newblock {\em {Cryogenic read-out system and resonator optimisation for the
  microwave SQUID multiplexer within the ECHo experiment}}.
\newblock {PhD thesis}, Kirchhoff-Institut f{\"u}r Physik, Universit{\"a}t
  Heidelberg, 7 2022.

\bibitem{Ho_Au_HC}
M.~Herbst, A.~Reifenberger, C.~Velte, H.~Dorrer, C.~E. D{\"u}llmann, C.~Enss,
  A.~Fleischmann, L.~Gastaldo, S.~Kempf, T.~Kieck, U.~K{\"o}ster,
  F.~Mantegazzini, and K.~Wendt.
\newblock Specific heat of holmium in gold and silver at low temperatures.
\newblock {\em J.Low Temp. Phys.}, Oct 2020.

\bibitem{Gamer2017}
L.~Gamer, C.~E. D{\"u}llmann, C.~Enss, A.~Fleischmann, L.~Gastaldo, C.~Hassel,
  S.~Kempf, T.~Kieck, and K.~Wendt.
\newblock Simulation and optimization of the implantation of holmium atoms into
  metallic magnetic microcalorimeters for neutrino mass determination
  experiments.
\newblock {\em Nucl. Instr. Meth. Phys. Res. A}, 854:139–148, 2017.

\bibitem{X-ray_data}
D.~Vaughan et~al. A.~C.~Thompson.
\newblock {Section 3.2 LOW-ENERGY ELECTRON RANGES IN MATTER}.
\newblock In {\em X-RAY DATA BOOKLET}. Lawrence Berkeley National Laboratory,
  University of California, Second edition, 2001.

\bibitem{NIS00}
NIST.
\newblock {X-Ray Mass Attenuation Coefficients}, checked in 2019.

\bibitem{Mantegazzini2021}
F.~Mantegazzini.
\newblock {\em {Development and characterisation of high-resolution metallic
  magnetic calorimeter arrays for the ECHo neutrino mass experiment}}.
\newblock {PhD thesis}, Kirchhoff-Institut f{\"u}r Physik, Universit{\"a}t
  Heidelberg, 7 2021.

\bibitem{Hengstler_dissertation}
D.~Hengstler.
\newblock {Development and characterization of two-dimensional metallic
  magnetic calorimeter arrays for the high-resolution X-ray spectroscopy}.
\newblock {\em Dissertation, Heidelberg University}, 2017.

\bibitem{readOut_paper}
F.~Mantegazzini and S.~Allgeier et~al.
\newblock Multichannel read-out for arrays of metallic magnetic calorimeters.
\newblock {\em Journal of Instrumentation}, 16(08):P08003, aug 2021.

\bibitem{Herbst_simulations}
Matthew Herbst, Arnulf Barth, Andreas Fleischmann, L.~Gastaldo, Daniel
  Hengstler, Neven Kovac, Federica Mantegazzini, Andreas Reifenberger, and
  Christian Enss.
\newblock Numerical calculation of the thermodynamic properties of silver
  erbium alloys for use in metallic magnetic calorimeters.
\newblock {\em Journal of Low Temperature Physics}, pages 1--9, 05 2022.

\bibitem{K_alpha_Hoelzer1997}
G.~Hoelzer, M.~Fritsch, M.~Deutsch, J.~Haertwig, and E.~Foerster.
\newblock {K$\alpha_{1,2}$ and K$\beta_{1,3}$ x-ray emission lines of the 3d
  transition metals}.
\newblock {\em Phys. Rev. A}, 56:4554, 1997.

\bibitem{enss_2005}
C.~Enss and S.~Hunklinger.
\newblock {\em Low-Temperature Physics}.
\newblock SpringerLink: Springer e-Books. Springer Berlin Heidelberg, 2005.

\bibitem{SQUID_handbook}
J.~Clarke and A.I. Braginski.
\newblock {\em The SQUID Handbook: Fundamentals and Technology of SQUIDs and
  SQUID Systems}.
\newblock Wiley, 2006.

\end{thebibliography}

\end{document}